\documentclass[11pt]{article}

\usepackage[final]{acl}
\usepackage{    tikz}
\usetikzlibrary{positioning, fit, shapes.multipart, arrows.meta, backgrounds}

\usepackage{times}
\usepackage{float}
\usepackage{latexsym}
\usepackage{booktabs}
\usepackage{enumitem}
\usepackage{amsmath}
\usepackage{placeins}

\usepackage[T1]{fontenc}

\usepackage[utf8]{inputenc}

\usepackage{microtype}

\usepackage{inconsolata}

\usepackage{graphicx}
\usepackage{xcolor}

\newcommand{\chunkbox}[1]{%
  \par\noindent
  \fcolorbox{red}{white}{%
    \parbox{\dimexpr\linewidth-2\fboxsep-2\fboxrule\relax}{%
      \raggedright\strut #1\strut
    }%
  }%
  \par\vspace{0.6em}%
}

\title{
\includegraphics[width=\textwidth]{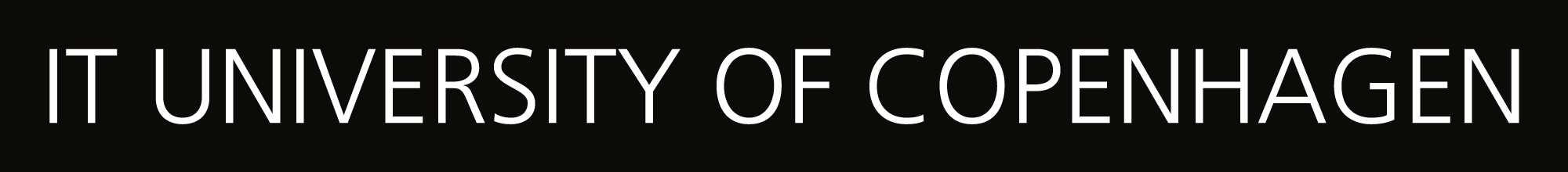}\\[4cm]

\Huge CRAwLeR - Cross-Reference Aware Legal Retrieval\\[1cm]}

\author{
  Maciej Jalocha\thanks{~Equal contribution.} \\
  IT University of Copenhagen \\
  \texttt{macja@itu.dk} \\
  \And
  William Michelsen\footnotemark[1]\thanks{~Contact person.} \\
  IT University of Copenhagen \\
  \texttt{wimi@itu.dk} \\
}

\begin{document}
\maketitle

\begin{center}
    \onecolumn
    \tableofcontents
    \twocolumn
\end{center}

\clearpage
\begin{abstract}
Existing benchmarks for context-aware chunk retrieval rely heavily
on repurposed task items and rarely demonstrate that their queries
genuinely require context, making score interpretation difficult.
We focus on a specific kind of context
dependence, legal cross-references, and introduce CRAwLeR, an
operationalization of a narrow, well-defined phenomenon: cross-reference-aware context utilization for
chunk retrieval in legal documents. Our pipeline detects legal
cross-references, identifies query candidates, links target chunks
to their relevant context, generates context-demanding queries with
an LLM, and filters them through both an adversarial non-contextual
baseline and an assurance prompt. We release CRAwLeR-DK and
CRAwLeR-PL, Danish and Polish datasets built with this
pipeline,\footnote{\url{https://github.com/pltier/crawler}}
alongside a strong Anthropic-style contextualization baseline.
Manual analysis finds that approximately 80\% of randomly sampled
queries genuinely target the labelled target chunk and require
context, with failures following systematic and named patterns.
The benchmarks are hard but not solved: best Recall@10 reaches
55\% on CRAwLeR-DK and 59\% on CRAwLeR-PL. Ablation and failure
analysis attribute the remaining gap to the contextualising LLM,
not the retriever. Even when the target is retrieved in the top
ten, labelled context chunks routinely outrank it. We are the first dataset for context-aware chunk retrieval to carefully consider construct validity and inspect our results in the light of such a narrow, well-defined phenomenon.
\end{abstract}


\section{Introduction}

The context windows of generative Large Language Models (LLMs) have grown enormously. This has driven the rise of the term Long-Context Language Models (LCLMs) \citep{liu2025comprehensivesurveylongcontext}. In parallel, "retrieval" increasingly refers to a generative model's ability to find the relevant parts of its own input, rather than to a separate retrieval system \citep{yang2024retrievalholisticunderstandingdolce}. Some work has gone further, exploring whether LCLMs can replace traditional retrieval pipelines altogether by ingesting entire corpora and outputting document or chunk identifiers directly \citep{lee2024longcontextlanguagemodelssubsume}. Despite these developments, more traditional retrieval tasks and solutions remain relevant.

Dual encoder setups are incredibly cost-effective and efficient \citep{Choi_2021}. RAG pipelines are also cheaper than approaches that dump full corpora into the long-context window of a generative model \citep{li-etal-2024-retrieval}. And in domains where provenance matters, retrieval will likely continue to play a key role even as generative answer quality improves \citep{Yang_2021, wang-etal-2026-autobool}. At the same time, embedders themselves have gained longer context windows, allowing them to encode longer documents, demanding relevant benchmarks \citep{10.5555/3692070.3693819, zhu-etal-2024-longembed}

\begin{figure*}[htbp]
    \centering
    \includegraphics[width=1\textwidth]{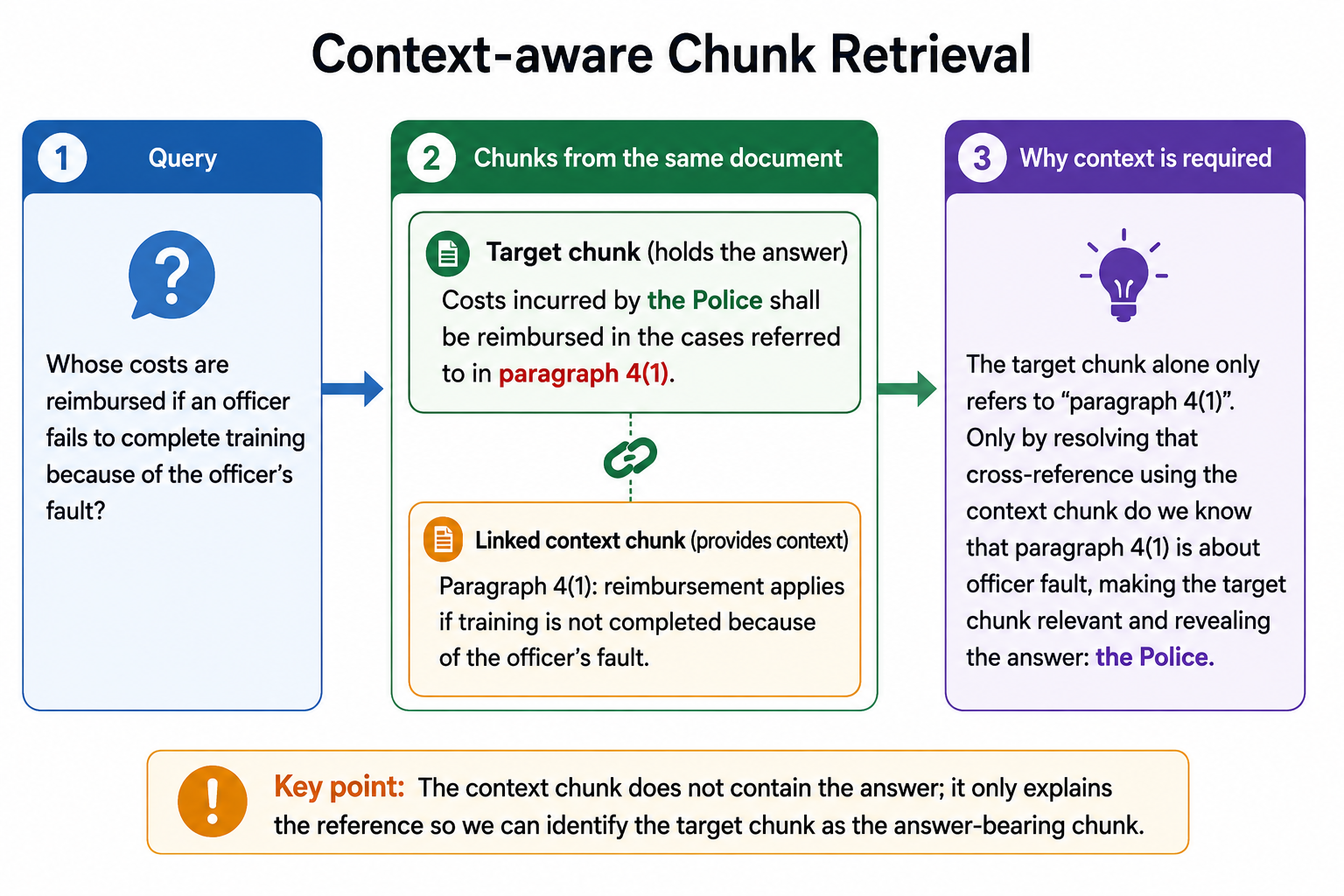}
    \caption{Visual definition and explanation of a possible context-aware chunk retrieval task item}
    \label{fig:context-aware-figure}
\end{figure*}

In practice, however, documents are usually segmented into chunks, and those chunks are encoded independently. This causes problems when a chunk's meaning is not self-contained \cite{conteb}. Without the surrounding context, retrieval accuracy can drop. Injecting chunks with the right context is therefore important. This makes context-aware chunk retrieval an interesting area of work. This broad task is illustrated in Figure~\ref{fig:context-aware-figure}. Many methods have been proposed, ranging from context-aware document segmentation \citep{chen-etal-2024-dense, duarte-etal-2024-lumberchunker} to text-level enrichment of chunks using generative LMs \citep{anthropiccontextualization}. Despite this, very few benchmarks exist to measure the capability. The ones that do exist are worth examining carefully, since they shape what claims can be made about how well models and methods perform on this sort of task. 

A recurring pattern in these benchmarks \citep{conteb, sitemb, dapr} is heavy reliance on repurposed task items. This is not a problem by itself. But it raises  validation concerns. If a dataset was not built to test context use, we need evidence that its items actually require context. We also need to know what kind of context is being tested, since context dependencies between passages of a document may take many forms.

We find that these benchmarks have problems with construct validity, understood in the sense given by \citet{messick1995validity}, as the degree to which evidence and theory support the interpretation of test scores for a proposed use or phenomenon. They repurpose existing task items without clearly justifying how those items measure context-aware chunk retrieval. In relation to this, they also have problems with scope. They aim to measure the broad capability of context-aware chunk retrieval, but, as said, context dependencies may come in many forms, and the benchmarks rarely specify which form they actually test. As a result, their scores become difficult to interpret. These benchmarks also leave methodological gaps that our work aims to address.

Given these limitations, we choose not to target the broad phenomenon of context-aware chunk retrieval directly. Instead, we have a smaller focus on a more specific sub-phenomenon that we define. This is cross-reference-aware context utilization for chunk retrieval in legal documents. We consider this to be a retriever's ability, given a query and a corpus of chunked legal documents, to rank the correct target chunk highly when its relevance can only be established by using information from one or more context chunks linked to the target by explicit legal cross-references. Figure~\ref{fig:context-aware-figure} provides a good illustration of this as well. Here, legal cross-references are understood as references between legal passages or legal texts in the sense of \citet{sannier2017automated}.

Narrowing in this way addresses some of the construct validity risks of prior work. There is less need to repurpose items, since legal cross-references are structurally explicit and automatically detectable \citep{sannier2017automated}, allowing for easier creation of task items. The scope problem is also reduced, since cross-reference resolution is a single, identifiable form of context dependency. The legal domain makes this feasible in two further ways. First, since legal cross-references are automatically detectable, it lets us identify good query candidates. This addresses a gap where prior pipelines could not effectively tell whether a chunk was suitable for query generation. This led to less efficient benchmark creation pipelines. Second, they are directly resolvable to specific context chunks, which lets us automatically store the relevant context for each target chunk. This addresses a gap where prior benchmarks had to feed entire documents to a generative model with no way to surface the relevant context for review.

To operationalize this task, we apply a non-contextual baseline as a filtering step. We leverage the structurally explicit cross-references to enable automated query generation and direct storage of the relevant context for each target chunk. We manually review query quality and also experiment with a synthetic quality assurance pipeline. Where relevant, we explain how each methodological choice contributes to construct validity, with a follow-up discussion in the results section.

This thesis makes 8 contributions. First, we give a critical review of existing context-aware chunk retrieval benchmarks, with a focus on how they define and validate context use. Second, we compare contextualization methods, including a recent extension to the method Late Chunking \citep{latechunking} called InSeNT \citep{conteb}, with Anthropic-style contextual retrieval \citep{anthropiccontextualization}. This comparison motivates Anthropic-style contextual retrieval as the main strong baseline we use for our results. Third, we operationalize cross-reference-aware context utilization as a specific retrieval task called CRAwLeR. Fourth, we introduce a pipeline that uses legal cross-references to build context-dependent retrieval items. Fifth, we present CRAwLeR-DK and CRAwLeR-PL, a Danish and Polish dataset based on CRAwLeR. Sixth, we evaluate contextual retrieval methods on these datasets and discuss how far the benchmark supports the claims made from its scores. Seventh, we introduce and evaluate a sliding-window variant of Anthropic-style contextual retrieval, where chunks are contextualised using overlapping local document windows rather than only the full document, motivated by common failure modes. Lastly, we perform an ablation study for Anthropic-style contextual retrieval.

\section{Literature Review}

\subsection{State of the Retrieval Framework}

\paragraph{Long-context models as retrievers.} Models that would be categorized as LCLMs are theoretically capable of solving the task in Figure~\ref{fig:context-aware-figure} too, for instance by being prompted to output document/chunk ids. However, this assumes the relevant collection of text can fit in the context \citep{xu2026surveymodelarchitecturesinformation, lee2024longcontextlanguagemodelssubsume}. This has made them emerge as a new paradigm for information retrieval \citep{seo-etal-2025-efficient}. However, problems such as information getting ‘lost-in-the-middle’ \citep{lostinthemiddle} and inefficient use of these large context windows where output quality may degrade as input size increases remain prevalent \citep{rulerwhatsrealcontext}. Other issues are the computationally expensive nature of using such models as redundant, unnecessary information will inherently end up being granted attention in the context window \citep{seo2025efficientlongcontextlanguage}. Furthermore, evaluations of LCLMs directly used for retrieval itself are still being conducted with relatively limited input sizes as real-world corpora, relevant for information retrieval, may scale far beyond even the largest context windows \citep{lee2024longcontextlanguagemodelssubsume}, leaving traditional pipelines viable as a result \citep{li-etal-2024-retrieval}. So, the remainder of our literature review, and thesis, will be framed primarily around considerations regarding models and solutions more commonly applied to Information Retrieval problems for instance, dense embedders and related solutions. This does not mean we completely ignore the potential of generative LMs for retrieval, as we will see in section \ref{sec:insent-vs-anthropic}.

\paragraph{Relevance of chunk retrieval.} Thanks to increasing context lengths, encoder-only architectures have become able to encode increasingly longer documents, and benchmarks have been created to measure this capability \citep{zhu-etal-2024-longembed, 10.5555/3692070.3693819}. But, retrieval of chunks is also highly relevant. Modern pipelines that involve retrieval often rely on some form of chunking, often called document segmentation. This is because many relevant documents exceed processing capacities of retrievers by their length \citep{wang-etal-2025-document}. Having more granularity may also facilitate the eventual generative LLM in RAG pipelines being able to more efficiently integrate retrieved information, rather than being overwhelmed by noise \citep{zhong-etal-2025-mix}. 

Chunking can also be made relevant from a more user-oriented perspective. One rationale is that often a document will be relevant for a given query only due to a small passage of information within that document. Computing the similarity between the whole document and the query may then underestimate the document’s relevance. Also, it may end up demanding unnecessary effort from the user to eventually extract what they are looking for, if the correct document is successfully retrieved \citep{documentrelevanceargument}. In an analysis conducted by \citet{dapr}, they found that many passages that users seek through Google queries are located, on average, 7.6 passages into the document. This significantly increases end-user effort if the whole document is returned rather than a precise chunk. In the legal domain, outputting full documents is common for retrieval setups. However, there has been a noted shift towards more fine-grained retrieval to support the more context-specific needs of legal professionals \citep{lexclipr}.  

\paragraph{Independent chunk encoding and context loss.} The problem, in regards to context use, is that documents are often chunked, and then the chunks are encoded independently. In settings where the queries are complex enough to demand context use to be solved, or in domains where chunks are rarely self-contained, this can become a problem for accurate retrieval \citep{conteb}. 

\subsection{Methods for Context-Aware Chunk Retrieval}
Many ways of improving the contextualization and document-awareness of chunks have been proposed. These range from solutions like boundary-aware chunking which aims to make chunks more self-contained or meaningful \citep{chen-etal-2024-dense, duarte-etal-2024-lumberchunker}, to changes to the embedding pipeline of long-context, dense embedding models, producing document-aware, contextual embeddings of chunks. 

\paragraph{Embedding level contextualization.} One example of the latter is Late Chunking, introduced by \citet{latechunking}. Instead of splitting a document into chunks and embedding each one independently, late chunking switches the order. The entire document is embedded first, and chunking is performed on the token-level embeddings after. Because every token mixes attention with the rest of the document, a chunk’s embedding includes information from the rest of the text. Then, mean-pooling is often applied. The approach is embedder agnostic; however, to be useful it requires longer context windows. \citet{latechunking} used jina-embeddings-v2
with its 8192-token window.


\paragraph{Text contextualization.} Another approach is index-time augmentation of the actual text using generative LMs. Contextual Retrieval, a method introduced by Anthropic, approaches the problem at the text level instead of the embedding level \citep{anthropiccontextualization}. Contextual Retrieval works by prepending a short, chunk-specific explanatory piece of information to each chunk before it is embedded. The same enriched text is also used to build a parallel BM25 \citep{BM25} index, yielding two sub-techniques the paper calls Contextual Embeddings and Contextual BM25. The pieces of information themselves are generated by an LLM (Claude 3 Haiku in the original write-up) that is prompted with the whole document plus the target chunk and asked for situating context, typically 50–100 tokens long. Prompt caching keeps this affordable, since the full document only needs to be loaded into the cache once per pass over its chunks. On Anthropic's evaluations across codebases, fiction, ArXiv and science papers, Contextual Embeddings alone cut the top-20 retrieval failure rate by 35\%, combining them with Contextual BM25 cut it by 49\%, and adding a reranking step on top brought the reduction to 67\%.  

\subsubsection{InSeNT vs. Anthropic-Style Contextualization}\label{sec:insent-vs-anthropic}
In this area of context-aware chunk retrieval, recent work by \citet{conteb} introduces InSeNT, a contrastive post-training method designed to improve contextual embeddings when combined with Late Chunking. The authors compare InSeNT with Anthropic-style contextualization and argue that the latter is inefficient. Although Anthropic-style contextualization reaches similar performance on their benchmark, they claim it does not scale well to large corpora because indexing is about 120 times slower: 1890.94 ms/doc, compared with 15.26 ms/doc for ModernBERT + InSeNT. We now discuss this claim, related aspects of the comparison, and some limitations of InSeNT that receive limited attention.

\paragraph{Scalability and amortization.} We see several issues with the scalability claim. First, the main additional cost of Anthropic Contextualization is paid at indexing time, not inference time. If the corpus is relatively stable, this cost can be amortized over the corpus lifetime. Second, Anthropic’s own write-up used one of its cheaper models, Claude 3 Haiku, while still reporting strong results \citep{anthropiccontextualization}. Third, the approach can be made cheaper through prompt caching, where attention states are reused across prompts that share overlapping document segments \citep{promptcaching}. The available evidence suggests, although it does not prove, that the main gains come from contextualizing chunks at all rather than from using expensive models to generate the context. For these reasons, we think the claim that Anthropic-style contextualization is “hardly scalable” to large corpora is too strong in general. The claim is more reasonable for corpora that change often, since changing documents require new chunk contextualizations. Or, for domains where accurate context generation requires slower and more expensive reasoning models.

\paragraph{Training versus indexing costs.} We also see oversights in the comparison between InSeNT and Anthropic Contextualization. Anthropic Contextualization is slower at indexing time, but this ignores the training and data-generation costs needed to produce an InSeNT model. \citet{conteb} do report some training cost for InSeNT, but they do not provide an end-to-end runtime comparison that includes the processing required to train the model. This matters because the two approaches shift costs to different places. Anthropic Contextualization pays its main cost when a corpus is indexed. InSeNT pays a larger cost before indexing, during data preparation and post-training. Both costs can be amortized: Anthropic Contextualization over the lifetime of a corpus, and InSeNT over the lifetime of the trained embedding model. As a result, Anthropic Contextualization may be simpler for relatively static corpora, including when used as a strong dataset baseline, while InSeNT may be more useful when the same trained model can be reused across many corpora.

\paragraph{Implementation complexity and model dependence.} We also argue that Anthropic-style contextualization is easier to understand and set up. It requires a relatively simple pipeline and limited prompt tuning. InSeNT is more involved. It requires long-document training data with queries mapped to chunks, and it shifts part of the work from indexing to training. \citet{conteb} had to create queries and repurpose existing data before applying their training method. InSeNT also adds another hyperparameter, $\lambda$ seq. As the authors note, the optimal value of this parameter depends on the task. This weakens the reusability of the resulting model across tasks, and potentially across corpora with different retrieval characteristics. We also note that we were unable to reproduce the reported ModernBERT + InSeNT results on ConTeb, while we were able to reproduce the Anthropic Contextual results.

Both approaches are model agnostic to some extent, but in different ways. Anthropic Contextualization is effectively agnostic to the model used for final retrieval. It changes the text itself before indexing, so the enriched chunks can be used with dense embedding models, BM25, or other text-based retrieval methods. This is reflected in Anthropic’s distinction between Contextual Embeddings and Contextual BM25. Anthropic also reported improvements across all tested embedding-model combinations and for Contextual BM25, although the gains varied across methods. 

InSeNT is less practically model agnostic. In principle, it can be applied to any embedding model with an 8192-token window that supports Late Chunking, exposes token states, and allows the implementation of InSeNT’s post-training objectives. In practice, this is a narrower requirement. If the embedding model is changed, the InSeNT training cost must be paid again. With Anthropic Contextualization, the main contextualization cost only needs to be repeated if the model used to generate the contextual text changes. In both cases, changing the retrieval model still requires recomputing the relevant representations.

\paragraph{Anthropic-style contextualization as a strong baseline.} Based on this comparison, we use Anthropic-style contextualization to establish strong baselines for CRAwLeR-DK and CRAwLeR-PL. Although it is slower at indexing time than InSeNT, its main cost is paid once and can be amortized for relatively stable corpora. It is also simple to implement, works with both dense and lexical retrievers, and does not require task-specific post-training. This makes it practical and transparent as a strong baseline.

\subsection{Benchmarks for Context-Aware Chunk Retrieval}
We now step back from individual methods and look across the prior work that studies context-aware chunk retrieval through the creation of benchmarks and/or datasets. The goal is to identify both limitations in these existing benchmarks and gaps that remain underexplored.

To the best of our knowledge, only a small body of work explicitly tries to build datasets or benchmarks to measure context utilization for chunk retrieval. A much broader set of papers becomes relevant if we include work where the authors merely note that context may help on their dataset because of properties of the corpus or domain. Here, we focus mainly on the stricter set. If a benchmark claims to measure context utilization, its value depends on whether its scores actually reflect that capability. This makes its methodology, results, and limitations especially important. Three works fit this scope: DAPR \citep{dapr}, ConTeb \citep{conteb}, and SitEmb \citep{sitemb}. They describe their benchmarks as measuring, respectively, “document aware passage retrieval,” “[models’] ability to leverage document-wide context,” and “situated retrieval capabilities.”

\paragraph{Repurposed-item pipelines.} These benchmarks are mainly built by repurposing task items from existing datasets, most often from question-answering (QA) or information retrieval datasets. The basic pipeline is similar across the three works. Long documents are split into chunks, either by using the original paragraph structure, as DAPR does for most of its datasets, or by applying a structure-aware chunker, as in ConTeb. Queries are then linked to chunks by reusing existing annotations. In some cases, an LLM is used to label which chunk in the gold document contains the answer. This is especially relevant for repurposed QA datasets where the question is not already tied to an annotated passage or chunk. Items are then filtered out when no suitable answer chunk can be found, since such items cannot function as retrieval items. SitEmb has the simplest pipeline. It inherits exact query-chunk pairs from PlotRetrieval and mainly repurposes the dataset at the document level, ensuring that documents exceed 100k tokens for RAG-related reasons. Overall, these benchmarks depend heavily on the validity of the original datasets they reuse.

\paragraph{Deliberate context-demanding construction.} ConTeb explicitly acknowledges that it is difficult to automatically and efficiently create items that genuinely require context use. Its authors describe “a fully automated and scalable method for generating high quality queries that effectively induce non-trivial context utilization” as an open challenge. DAPR and ConTeb do more deliberate work on this problem in some parts of their benchmarks. DAPR manually analyzes queries from its constituent datasets to decide whether they require context use and, if so, what kind. But this analysis is only performed for a subset of queries from one of its five datasets, as they name a full manual review too time-consuming. ConTeb’s “controlled settings” take a different approach. These datasets are either created by manually writing context-demanding queries or by giving an entire document to a generative LLM, asking it to write a query for a specific chunk while also injecting ambiguity into that chunk so that the chunk is not self-sufficient.

\subsubsection{Construct-Validity Concerns in Existing Benchmarks}
Because these benchmarks rely so strongly on repurposed items, and because they only briefly explain how their construction captures context utilization, construct validity is a useful lens for evaluating them. Construct validity is complex, but for our purposes we define it in line with the work by \citet{constructvaliditymetasurvey}. That is, the extent to which scores on a dataset or benchmark provide evidence for the phenomenon the benchmark claims to measure. Here, the phenomenon is context-awareness. Or, more precisely, a model’s ability to use document context for chunk retrieval. We identify two main construct-validity concerns in the prior work: how task items are sampled, and how this target phenomenon is scoped.

\paragraph{Sampling.} The first concern is sampling. These benchmarks appear to rely on a mix of convenience sampling  and criterion sampling. Convenience sampling means selecting task items mainly because they are available \citep{conveniencesampling}. Criterion sampling means filtering those items according to one or more conditions \citep{criterionsampling}. This is not automatically a problem, but especially convenience sampling creates risks for construct validity \citep{constructvaliditymetasurvey}. By relying on convenience sampling, the benchmarks also rely on the original datasets to capture the phenomenon they now want to measure: context-awareness or context utilization for chunk retrieval. With a few exceptions, and from what we can tell, the original datasets were not designed to specifically measure context utilization or a closely related reasoning capability. We are not claiming that context use is irrelevant to those original items. Rather, because context use was not a direct design target, it is unclear how many items genuinely require it, or how strongly they require it.

\paragraph{Criterion filters and item validation.} The criterion sampling side does not fully solve this problem. Most filters applied to repurposed items ensure that the items work as retrieval items. They do not ensure that the items require context utilization. ConTeb does apply a method to make sure that a chunk is not self-sufficient given a query. However, this is not the same as proving that document context is required. A chunk can fail to answer a query on its own for several reasons: the query may be underspecified, the answer may be missing, or the item may be noisy. Showing that the chunk is insufficient is therefore weaker than showing that the intended surrounding context is necessary.

There is also little explicit discussion, explanation, or justification of construct validity. Recent recommendations for LLM benchmark design call for authors to discuss and consider construct validity directly, including the strengths and weaknesses of reused datasets and the effects of modifying them \citep{constructvaliditymetasurvey, whatwillittake, wideworldbenchmark}. At best, these works engage with such recommendations only briefly. More importantly, there is no manual or synthetic review of whether the final task items, or a representative sample of them, actually require context utilization. Or, a more careful analysis of failure modes to determine if other phenomena are playing a meaningful role. ConTeb also does not report a review of whether the outputs of its synthetic controlled pipeline truly demand context utilization, or to what extent. The only exceptions are the manually labeled subset of DAPR and the manually written subset of ConTeb.

A relatively simple, though imperfect, mitigation would be adversarial filtering \citep{whatwillittake}: removing items that a non-contextual baseline can solve with high confidence. None of the benchmarks apply such a step. 

Our claim is therefore twofold. First, we are doubtful that their sampling approach establishes construct validity. Second, even if many items do require context, this is not shown transparently. In both cases, it becomes unclear how far scores on these benchmarks can be interpreted as performance on context-demanding chunk retrieval. This problem is made worse by how the benchmarks define the scope of the task.

\paragraph{Scope of context utilization.} The second construct-validity concern is scope. Context utilization can take many forms. Dependencies between chunks of text may involve coreference resolution \citep{coreference}, topic structure \citep{topicunderstanding}, discourse relations \citep{discourserelation}, bridging anaphora \citep{bridging}, and other concepts or phenomena. To the best of our knowledge, there is no complete, agreed-upon taxonomy. There is also no clear evidence that a method able to use one kind of context will necessarily use another kind well. This matters because the type of context being measured directly affects what a benchmark score can mean. A recent survey on construct validity in LLM benchmark design therefore recommends defining the measured phenomenon carefully and explaining which aspects are included or excluded \citep{constructvaliditymetasurvey}. The benchmarks discussed here mostly do not do this.

\paragraph{How these benchmarks scope context use.} DAPR labels a subset of its benchmark with clear categories: coreference resolution, main topic, multi-hop reasoning, and acronym resolution. These labels include definitions and examples, based on the manual analysis discussed above. But the labeling only applies to that subset. Most of the benchmark has no such categorization. 

ConTeb labels its whole benchmark by the type of context utilization required in each constituent dataset, but it does not define those labels. For some labels, this is less problematic because their meaning is well established in NLP, such as coreference resolution. For others, such as “document-reasoning,” several plausible meanings exist. Without a definition, it is unclear how performance under that label should be interpreted, which weakens construct validity.

SitEmb provides neither categories nor definitions. It says the task requires “context-aware embedding capability” and that the original repurposed dataset measures “semantic associations.” But semantic association can be a vague term in the world of NLP. In our view, SitEmb does not define the term well enough for this task. PlotRetrieval \citep{plotretrieval}, the original dataset, does define semantic association more clearly and introduces five categories. But neither PlotRetrieval nor SitEmb clearly explains how those categories relate to context-aware chunk retrieval. Instead, SitEmb relies on prior work showing that a context-aware method improves results on a repurposed version of PlotRetrieval \citep{sitembpriorwork}. We think it is suggestive evidence, but not enough. Also, it still does not tell us what kind of context utilization the task demands.

\paragraph{Implication for this thesis.} A key part of establishing construct validity is that the phenomenon being operationalized must be defined carefully \citep{messick1995validity}. These papers do define the broad task of context-aware chunk retrieval reasonably well. But their definitions of the relevant subcomponents are incomplete or missing. Combined with the sampling problem, this means that even when items genuinely require context, the score does not tell us what kind of context-use capability is being assessed.
To summarize, strong construct validity requires a clear chain of reasoning from the definition of the phenomenon, to its operationalization as a task, to the selection of concrete task items, to the implementation of the benchmark, and finally to the claims made from its scores \citep{constructvaliditymetasurvey}. Based on this, our work will focus on making the target form of context utilization explicit, explaining how our task items measure it, and discussing where the resulting benchmark succeeds or fails in supporting validity claims.

\subsubsection{Methodological Gaps in Automatic Query Generation}
\paragraph{Full-document synthetic generation.} Setting construct validity aside, ConTeb’s synthetic pipeline also raises methodological issues relevant to our work. In ConTeb’s controlled setting, the full relevant document is given to a generative LLM twice: once to rephrase a target chunk and inject ambiguity, and once to write the query for that chunk. Because the whole document must be included in two separate prompts, this approach can quickly become expensive for large corpora of long documents. The ConTeb authors also seem to acknowledge this. Accuracy is another concern. As documents grow, the ratio of useful context signal to irrelevant text around the target chunk likely falls. This may make it harder for the model to inject the right ambiguity or to generate a query that truly requires context. Even when the generative model’s context window is large enough, prior work suggests that output quality can degrade at input lengths far below the maximum supported window \citep{fakecontextlengths}.

\paragraph{Candidate detection and context identification.} Based on our reading of ConTeb, two underlying issues limit the efficiency of such automatic query generation. First, the pipeline has no way to detect whether a paragraph is a good candidate for query generation. This creates a quality problem, but also a cost problem: computation may be spent trying to generate queries for paragraphs that are not suitable in the first place. Second, and more importantly, the pipeline has no way to identify which parts of the document provide the most relevant context for a given chunk. Without that information, giving the entire document to the generative model seems to be the most reasonable option.

\subsection{Legal Cross-References as an Operationalization Route} 
\paragraph{Cross-references as explicit context dependencies.} The legal domain offers a promising way to address these issues. Legal texts often contain references between documents and between passages within the same document. These are usually called legal cross-references. Such references can create context dependencies between passages, since “analysts often need to follow the cross references while looking for additional information” \citep{sannier2017automated}. This seems useful for our purposes because previous work has already successfully explored how to detect and resolve such references automatically. In other words, prior work by \citet{sannier2017automated} has studied how to identify the natural-language phrases that express a reference and link them to the passages they refer to. This indicates that the legal domain can allow us to find query candidates and identify the most relevant context for each target chunk.
Other work has also explored classifiers that characterize the semantic intent behind a legal cross-reference \citep{semanticintentclassification}. This may be useful to measure ones task space coverage and identify gaps in task items.

\paragraph{Other reference-based context dependencies.} To be clear, similar explicit references also exist outside law. For example, in scientific articles, work has studied automatic detection and linking of text and table-cell relationships \citep{10.1145/3242587.3242617}. Related forms of context dependency have also been studied without explicit references. In coreference resolution, some work has explored finding all event mentions that corefer with an event mentioned in a query \citep{eirew-etal-2022-cross}. In multi-hop reasoning, some work has studied how to answer multi-hop questions over long documents by iteratively attending to relevant document parts \citep{sun2021iterativehierarchicalattentionanswering}. Nonetheless, we find the legal domain to be the most promising, given how explicit and directly the context dependencies occur in the text.

\subsection{Summary of Gaps and Limitations} We can now state the main gaps and limitations identified in this literature review. We use the two terms deliberately. A major limitation is that the existing benchmarks generally do not use manual or synthetic review pipelines to verify that repurposed or newly created items actually depend on context use. Small, non-representative subsets of DAPR and ConTeb are exceptions. A second limitation is that the specific forms of context utilization being measured are often under-defined, under-operationalized, or not identified for most task items. This is paired with little discussion of whether the benchmark design establishes construct validity, especially when task items are repurposed from prior datasets.
On the methodological side, a major gap is the lack of an efficient, automated process for creating queries that demand non-trivial context utilization. Another gap is the absence of adversarial filtering with a non-contextual method or model. Such filtering could help remove items that do not actually require context. A further gap is that these benchmarks do not automatically detect and store the most relevant context for each target chunk. This is difficult, so we treat it as a gap rather than a limitation. But it could matter a lot. It could reduce query-generation cost by avoiding the need to provide an entire document to the generative model. It could also improve interpretability, support error analysis, and make it easier to review whether queries truly demand context. SitEmb is not a full counterexample here. It does define what counts as context more precisely than the other benchmarks, but it still does not give the same kind of directly stored, reviewable, most-relevant context chunk that would support pipeline inspection and model analysis. Lastly, all three benchmarks are in English, leaving a need for equivalent benchmarks in other languages. These gaps, and the promise of the legal domain, motivate a narrow, cross-reference-based task.

\section{Task Definition \& Data}

\subsection{Cross-reference-aware context utilization for chunk retrieval in legal documents}\label{sec:phenomenom-definition}

As established in our literature review, attempting to measure a broad or complex phenomenon, such as context utilization for chunk retrieval, carries two related risks. The first is construct underrepresentation, where the benchmark fails to capture all aspects of the phenomenon it claims to measure \citep{messick1995validity, benchmarkingepistemologyconstructvalidity}. The second is poor score interpretability, which follows from a lack of specificity in what is being measured \citep{wideworldbenchmark}. Based on the gaps identified in the literature review, the promise of the legal domain, and to mitigate both risks, we focus on a narrower phenomenon: cross-reference-aware context utilization for chunk retrieval in legal documents. Operationally, this is a retriever's ability, given a query and a corpus of chunked legal documents, to rank the correct target chunk highly when the target chunk's relevance can only be established by using information from one or more context chunks linked to the target chunk by explicit legal cross-references.
 
\subsection{The CRAwLeR Task}
We name our task to measure this phenomenon CRAwLeR, short for Cross-reference-aware Legal Retrieval:

Let $\mathcal{D}$ denote a corpus of legal documents. Each document $d \in \mathcal{D}$
is partitioned by a chunking function $P$ into a sequence of chunks,
$P(d) = \{c_1, c_2, \ldots, c_{N_d}\}$. The full chunk corpus is
$\mathcal{C} = \bigcup_{d \in \mathcal{D}} P(d)$. For each chunk $c \in \mathcal{C}$,
let $R(c) \subseteq \mathcal{C}$ denote the set of chunks that $c$ explicitly
references via legal cross-references; we refer to these as the
\emph{context chunks} of $c$.

A CRAwLeR task item is a pair $(q, c^*)$, where $q \in \mathcal{Q}$ is a query
and $c^* \in \mathcal{C}$ is its gold target chunk. Valid task items satisfy the
condition that the relevance of $c^*$ to $q$ cannot be established from the
content of $c^*$ alone, and requires information from at least one chunk in
$R(c^*)$.

Given an embedding function $\phi$ and a similarity function $f$, a standard
non-context-aware retrieval system computes
\[
    s(q, c) = f\bigl(\phi(q),\, \phi(c)\bigr),
\]
while a context-aware system should compute
\[
    s(q, c) = f\bigl(\phi(q),\, \phi(c, R(c))\bigr).
\]
In either case, the system returns the top-$K$ chunks ranked by $s$, with the
objective that $c^*$ be ranked as high as possible. The validity condition on
task items entails that the non-context-aware form should not, in general,
suffice to rank $c^*$ correctly. The concept of cross-reference resolution, as defined by \citet{sannier2017automated} is thus relevant for CRAwLeR: It means to take expressions that denote cross-references, and to interpret these expressions and link them to the referenced passage or chunk.

We construct two datasets under this task definition, one in Danish and one in Polish. We refer to them as CRAwLeR-DK and CRAwLeR-PL, respectively.

We want to be clear that, despite the narrower scope of our task, we will still consider and examine our work through the lens of construct validity. The framework of construct validity is most often applied to broad, latent, 'theoretical constructs', such as 'reasoning' \citep{mcgrath2005conceptual, constructvaliditymetasurvey}. It remains useful, however, for thinking about how scores should be interpreted in general. We follow the recommendation of prior work not to dismiss construct validity simply because the scope is narrower, or because the target is not necessarily a latent, 'theoretical construct' \citep{messick1990validity}. Concretely, and in line with recommendations on NLP benchmark construction \citep{whatwillittake, constructvaliditymetasurvey}, we will explain in the methodology, where relevant, how each design choice contributes to measuring the phenomenon. We will also conduct an analysis of errors and predictions, and revisit construct validity in the discussion section. To be clear, this means that scores on CRAwLeR should be interpreted as evidence about this specific facet of context utilization (cross-reference-aware context utilization for chunk retrieval in legal documents), and not as a general measure of context-aware chunk retrieval as a whole. 

\subsection{Data Sources}

Based on this, we will use documents from RetsInformation~\footnote{https://www.retsinformation.dk} and ELI API for Polish Acts~\footnote{https://api.sejm.gov.pl/eli.html}. There are a number of reasons we find these sources suitable. They are public domain and easily accessible which should facilitate future work on our datasets. Secondly, the passages of the documents contain plenty of direct within-document cross references. This allows us to automatically detect query candidates and referenced context chunks as we desire. The details, including the sizes, and source URLs, of the datasets can be found in the Appendix~\ref{app:data-sources}.

\section{Methodology}

\subsection{Selection of Documents}
We pick documents that are at least 32k tokens long. This threshold is motivated by the idea to make it more difficult to apply Late Chunking approaches, which are limited by embedder max input size. The embedding architectures, we found, scaled up to 32K tokens long input at most \citep{zhu-etal-2024-longembed, saadfalcon2024benchmarkingbuildinglongcontextretrieval}.

While we are not evaluating Late Chunking approaches, this is a consideration for future researchers that might use the datasets.

There were not many documents in the data sources passing this threshold, and that is why we ended up with only 7 per dataset.

In order to increase the retrieval difficulty we pick documents from similar domains. For each dataset, we create two clusters of the documents, being similar semantically and topic-wise. (Similarity heatmaps in Appendix~\ref{app:document-similarity-heatmaps}) The 7th document has no cluster, acting as an distractor-free baseline.  Had the documents been picked randomly, the chunks would have most likely been more distinguishable, reducing the datasets' difficulty.

\subsection{Chunking Approach}\label{sec:appendix}
Figure~\ref{fig:polish-chunking} shows chunking strategy for the CRAwLeR-PL dataset. A chunk is meant to capture an individual piece of information. In practice it was split by newlines.
\begin{figure}[ht]
  \centering
  \setlength{\fboxsep}{6pt}
  \begin{minipage}{0.48\textwidth}
    \small\ttfamily
    \chunkbox{Art. 37s. 1. Head of the organizational unit of the State Fire Service referred to in Article 37l point 2, in relation to firefighters delegated to perform tasks in civilian institutions:}
    \chunkbox{\hspace*{2em}1) is a disciplinary superior and supervises compliance with official discipline, with the exception of secondment to the Office of Internal Oversight Services;}
    \chunkbox{\hspace*{2em}2) makes decisions on matters arising from the service relationship, in accordance with the principles specified in the Act.}
    \chunkbox{2. The provision of paragraph 1 does not affect the powers of the head of the civil institution in which the firefighter performs tasks.}
  \end{minipage}
  \caption{Chunking applied in the CRAwLeR-PL dataset, on a translated example.}
  \label{fig:polish-chunking}
\end{figure}

Figure~\ref{fig:danish-chunking} shows how chunking was performed on the CRAwLeR-DK dataset. The splits capture "pieces" (stykke / stk. These are parts of legal section. Legal sections being denoted by the symbol §).

\begin{figure}[ht]
  \centering
  \setlength{\fboxsep}{6pt}
  \begin{minipage}{0.48\textwidth}
    \small\ttfamily
    \chunkbox{SECTION 12. A public housing association's articles of association must contain provisions on:\\
    \hspace*{2em}1) Name and domicile of the housing association.\\
    \hspace*{2em}2) Capital relations of the housing association.}
    \chunkbox{Pcs. 2. Articles of association for a guarantee organization must also contain provisions on holding a guarantor meeting and convening it.}
  \end{minipage}
  \caption{Chunking applied in the CRAwLeR-DK dataset, on a translated example.}
  \label{fig:danish-chunking}
\end{figure}

\FloatBarrier

\begin{figure*}[t]
\centering
\resizebox{\textwidth}{!}{%
\begin{tikzpicture}[remember picture,
    mainbox/.style={draw, rectangle, solid, thick, align=left, text width=12cm, inner sep=8pt, fill=blue!10},
    mainimplicit/.style={draw=gray, rectangle, dashed, thin, align=left, text width=12cm, inner sep=8pt, fill=gray!5, text=darkgray},
    contextbox/.style={draw=black, rectangle, solid, thick, align=left, text width=12cm, inner sep=8pt, fill=green!10},
    implicitbox/.style={draw=gray, rectangle, dashed, thin, align=left, text width=12cm, inner sep=8pt, fill=gray!5, text=darkgray},
    inlinebox/.style={draw, rectangle, solid, thin, inner sep=2pt, fill=orange!20},
    arrow/.style={->, >=Stealth, thick, color=orange!80!black},
    dots/.style={align=center, font=\bfseries\Large},
    querybox/.style={draw=purple, rectangle, solid, line width=1.5pt, align=left, text width=10cm, inner sep=18pt, fill=purple!5, font=\Large},
    qaddressed/.style={draw=black, rectangle, solid, thick, inner sep=3pt, fill=green!10, font=\Large, rounded corners=2pt},
    qunaddressed/.style={draw=gray, rectangle, dashed, thick, inner sep=3pt, fill=gray!5, font=\Large, rounded corners=2pt, text=darkgray},
    queryarrow/.style={->, >=Stealth, ultra thick, dashed, color=purple},
    targetarrow/.style={->, >=Stealth, line width=1.5mm, solid, color=purple!80!black},
    genbox/.style={draw=black!80, rectangle, solid, ultra thick, align=center, inner sep=15pt, fill=yellow!20, font=\Large\bfseries, rounded corners=8pt},
    spinearrow/.style={->, >=Stealth, line width=2mm, draw=black!70},
    innerdots/.style={align=center, font=\bfseries\Large, fill=blue!2.5, text=black!80, rounded corners=4pt, inner sep=4pt}
]
%
%
\node[implicitbox] (contextTop1) at (0,0) {
    \textbf{(Implicit Context of a Context Chunk)}\\
    4. The agreement referred to in paragraph 2 sets out the mutual rights and obligations of the parties related to the referral... in the case of:
};
\node[innerdots, below=0.2cm of contextTop1] (dotsTop1) {[...]};
\node[contextbox, below=0.2cm of dotsTop1] (contextTop2) {
    \textbf{(Context Chunk)}\\
    1) failure to complete training, school, higher or postgraduate studies at home or abroad or legal training as a result of obtaining a negative final grade or discontinuing studies due to the fault of a police officer;
};
%
\node[innerdots, below=0.6cm of contextTop2] (dotsMid1) {[...]};
\node[mainimplicit, below=0.2cm of dotsMid1] (implicitTarget) {
    \textbf{(Implicit Context of a Target Chunk)}\\
    5. Reimbursement of costs incurred shall be in the amount of
};
\node[innerdots, below=0.2cm of implicitTarget] (dotsTarget) {[...]};
\node[mainbox, below=0.2cm of dotsTarget] (target) {
    \textbf{(Target Chunk)}\\
    1) costs incurred by the Police - in the cases referred to in \\
    {\tikz[remember picture, baseline=(p4.base)]\node[inlinebox](p4){[paragraph 4 point 1]};} and 
    {\tikz[remember picture, baseline=(art41.base)]\node[inlinebox](art41){[Article 41 paragraph 1 points 3-5]};} 
    and {\tikz[remember picture, baseline=(p2.base)]\node[inlinebox](p2){[paragraph 2 points 2 and 8]};}
};
%
\node[innerdots, below=0.6cm of target] (dotsMid2) {[...]};
\node[implicitbox, below=0.2cm of dotsMid2] (contextBottom1) {
    \textbf{(Implicit Context of a Context Chunk)}\\
    Art. 41. 1. A police officer is dismissed from service in cases of:
};
\node[innerdots, below=0.2cm of contextBottom1] (dotsBottom1) {[...]};
\node[contextbox, below=0.2cm of dotsBottom1] (contextBottom2) {
    \textbf{(Context Chunk)}\\
    3) imposing a disciplinary penalty of dismissal from service;
};
%
\node[fit=(contextTop1)(dotsTop1)(contextTop2)(dotsMid1)(implicitTarget)(dotsTarget)(target)(dotsMid2)(contextBottom1)(dotsBottom1)(contextBottom2), inner sep=35pt] (tempOval) {};

\node[dots, above=0.3cm of tempOval.north] (outDotsTop) {[...]};
\node[dots, below=0.3cm of tempOval.south] (outDotsBottom) {[...]};

\node[fit=(tempOval)(outDotsTop)(outDotsBottom), inner sep=30pt] (tempDoc) {};

\begin{scope}[on background layer]
    \node[draw=blue!40, thick, solid, rounded corners=8pt, fill=blue!2, fit=(tempDoc)] (docWrapper) {};
    \node[draw=orange!80, ultra thick, dashed, rounded corners=60pt, fill=orange!10, fit=(tempOval)] (ovalWrapper) {};
\end{scope}
%
\node[anchor=north west, font=\bfseries\huge, color=blue!80!black, xshift=15pt, yshift=-15pt] at (docWrapper.north west) {Document};
\node[anchor=north east, font=\bfseries\LARGE, color=orange!80!black, xshift=-25pt, yshift=-15pt] at (ovalWrapper.north east) {Prompt Payload};
%
\draw[arrow, overlay] (p4.north) to[out=90, in=270, looseness=1.5] node[pos=0.22, sloped, above, font=\small] {cross-reference} (contextTop2.south);
\draw[arrow, overlay] (art41.east) to[out=-10, in=90, looseness=1.5] node[pos=0.55, sloped, below, font=\small] {cross-reference} (contextBottom2.north);
%
%
\node[genbox, right=1cm of docWrapper.east] (generator) {Query\\Generation\\Pipeline};

%
%
\node[querybox, right=2.5cm of generator.east] (query) {
    \textbf{Generated Query:}\\[1.5ex]
    In cases where a police officer has \\
    {\tikz[remember picture, baseline=(q1.base)]\node[qaddressed](q1){not completed training};}, has been \\
    {\tikz[remember picture, baseline=(q2.base)]\node[qaddressed](q2){dismissed due to a disciplinary penalty};}, \\
    has been {\tikz[remember picture, baseline=(q3.base)]\node[qunaddressed](q3){convicted by a final court judgment};} \\
    or has {\tikz[remember picture, baseline=(q4.base)]\node[qunaddressed](q4){renounced his Polish citizenship};}, \\
    is there any reimbursement of costs incurred\\by the Police?
};
%
%
\draw[spinearrow] (ovalWrapper.east |- generator.west) 
    -- node[above, font=\LARGE\bfseries] {Input} 
    (generator.west);
\draw[spinearrow] (generator.east) 
    -- node[above, font=\LARGE\bfseries] {Outputs} 
    (query.west |- generator.east);
%
%
\draw[targetarrow, overlay] (query.north west) 
    to[out=120, in=80, looseness=1] node[pos=0.5, sloped, above, font=\LARGE\bfseries] {targets} 
    (target.north east);
%
\draw[queryarrow, overlay] (q1.west)
    to[out=160, in=0, looseness=0.9] node[pos=0.4, sloped, above, font=\Large] {derived from} 
    (contextTop2.east);
%
\draw[queryarrow, overlay] (q2.west)
    to[out=200, in=0, looseness=0.9] node[pos=0.5, sloped, above, font=\Large] {derived from} 
    (contextBottom2.east);
\end{tikzpicture}
}
\caption{Visualisation of the definitions of implicit context, context chunks, explicit references, and target chunk, and how these are input into Query Generation Pipeline. Polish translated example used as an example.}
\label{fig:context-chunking}
\end{figure*}
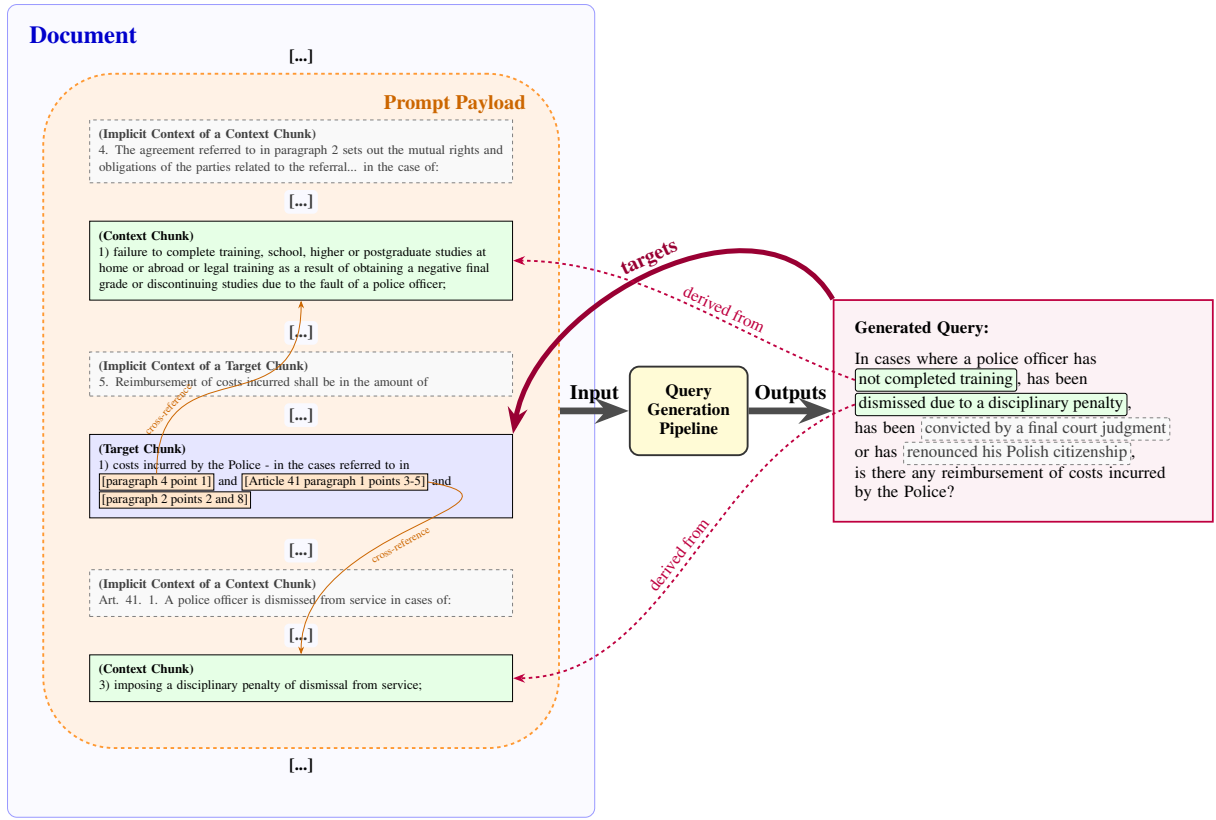

\subsection{Definitions of Chunk Types}
To faciliate creation and understanding of CRAwLeR task items, we define some important classes of chunks:
\begin{itemize}
    \item Target chunk: This chunk holds the answer to a query and is labeled as such. However, one can only know that the target chunk is the answer to its corresponding query by resolving its legal cross-references. Understanding of relevant implicit context chunks may also provide clues.
    \item Context Chunk(s): For a given target chunk, these are chunks referenced by the cross-references in the target chunk. I.e. resolving the target chunk's cross-reference leads to identification of these chunks and their information.
    \item Implicit Context of Target Chunk: In CRAwLeR-DK, these are the target chunk's neighboring chunks in the same legal section §. In CRAwLeR-PL, these are chunks in the same legal section as the target chunk, denoted by 'Art.', that may say something about the target chunk. 
    \item Implicit Context of Context Chunk(s): Same definition as previous point, but applied to context chunks.
    \item Query Candidate: A chunk that contains an explicit legal cross-reference to another chunk. This suggests a context demanding query can be crafted with respect to the chunk. It's worth defining this, so we avoid spending compute trying to craft queries for chunks that do not lend themselves to the CRAwLeR task.
\end{itemize}
These chunk types, and their relationships, are illustrated in figure \ref{fig:context-chunking}. These definitions are meant to facilitate crafting queries that measure cross-reference-aware context utilization for chunk retrieval in legal documents, as described in the next section.

\subsection{Query Generation Pipeline}
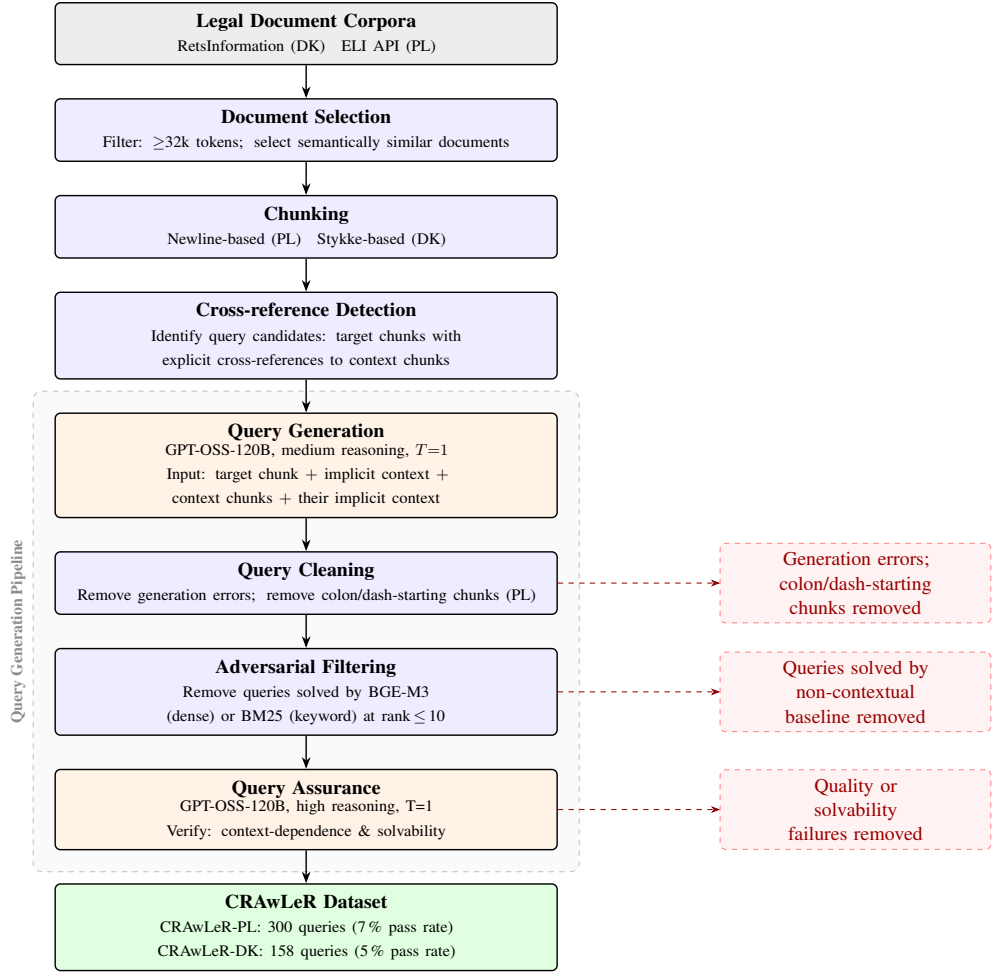
\begin{figure*}[t]
\centering
\resizebox{0.82\textwidth}{!}{%
\begin{tikzpicture}[
    node distance=0.65cm and 3.2cm,
    mainbox/.style={draw, rectangle, rounded corners=3pt, thick, align=center,
                    text width=9.5cm, minimum height=1.1cm, inner sep=6pt, fill=blue!7},
    inputbox/.style={draw, rectangle, rounded corners=3pt, thick, align=center,
                    text width=9.5cm, minimum height=1.1cm, inner sep=6pt, fill=gray!14},
    outputbox/.style={draw, rectangle, rounded corners=3pt, thick, align=center,
                    text width=9.5cm, minimum height=1.1cm, inner sep=6pt, fill=green!12},
    modelbox/.style={draw, rectangle, rounded corners=3pt, thick, align=center,
                    text width=9.5cm, minimum height=1.1cm, inner sep=6pt, fill=orange!10},
    rejectbox/.style={draw, rectangle, rounded corners=3pt, thin, dashed, align=center,
                    text width=5.0cm, minimum height=0.85cm, inner sep=5pt,
                    fill=red!5, draw=red!50, text=red!60!black},
    arrow/.style={->, >=Stealth, thick},
    rejectarrow/.style={->, >=Stealth, thin, dashed, color=red!55!black},
]

\node[inputbox] (docs)
    {\textbf{Legal Document Corpora}\\
     {\small RetsInformation (DK)\quad ELI API (PL)}};

\node[mainbox, below=of docs] (select)
    {\textbf{Document Selection}\\
     {\small Filter: $\geq$32k tokens;\; select semantically similar documents}};
\draw[arrow] (docs) -- (select);

\node[mainbox, below=of select] (chunk)
    {\textbf{Chunking}\\
     {\small Newline-based (PL)\quad Stykke-based (DK)}};
\draw[arrow] (select) -- (chunk);

\node[mainbox, below=of chunk] (xref)
    {\textbf{Cross-reference Detection}\\
     {\small Identify query candidates: target chunks with explicit cross-references to context chunks}};
\draw[arrow] (chunk) -- (xref);

\node[modelbox, below=of xref] (qgen)
    {\textbf{Query Generation}\\
     {\small GPT-OSS-120B, medium reasoning, $T{=}1$\\
      Input: target chunk $+$ implicit context $+$ context chunks $+$ their implicit context}};
\draw[arrow] (xref) -- (qgen);

\node[mainbox, below=of qgen] (qclean)
    {\textbf{Query Cleaning}\\
     {\small Remove generation errors;\; remove colon/dash-starting chunks (PL)}};
\draw[arrow] (qgen) -- (qclean);

\node[mainbox, below=of qclean] (adv)
    {\textbf{Adversarial Filtering}\\
     {\small Remove queries solved by BGE-M3 (dense) or BM25 (keyword) at rank\,$\leq$\,10}};
\draw[arrow] (qclean) -- (adv);

\node[modelbox, below=of adv] (qa)
    {\textbf{Query Assurance}\\
     {\small GPT-OSS-120B, high reasoning, T=1\\
      Verify: context-dependence \& solvability}};
\draw[arrow] (adv) -- (qa);

\node[outputbox, below=of qa] (final)
    {\textbf{CRAwLeR Dataset}\\
     {\small CRAwLeR-PL: 300 queries (7\,\% pass rate)\quad
             CRAwLeR-DK: 158 queries (5\,\% pass rate)}};
\draw[arrow] (qa) -- (final);

\node[rejectbox, right=3.2cm of qclean] (rej_clean)
    {Generation errors;\\colon/dash-starting\\chunks removed};
\node[rejectbox, right=3.2cm of adv] (rej_adv)
    {Queries solved by\\non-contextual\\baseline removed};
\node[rejectbox, right=3.2cm of qa] (rej_qa)
    {Quality or\\solvability\\failures removed};

\draw[rejectarrow] (qclean.east) -- (rej_clean.west);
\draw[rejectarrow] (adv.east)    -- (rej_adv.west);
\draw[rejectarrow] (qa.east)     -- (rej_qa.west);

\begin{scope}[on background layer]
    \node[draw=black!25, dashed, rounded corners=5pt, fill=black!2,
          fit=(qgen)(qclean)(adv)(qa),
          inner sep=12pt,
          label={[rotate=90, anchor=south, font=\small\bfseries,
                  text=black!45, inner sep=4pt]left:Query Generation Pipeline}
         ] {};
\end{scope}

\end{tikzpicture}
}
\caption{Overview of the CRAwLeR dataset creation pipeline. Legal documents are
chunked and cross-references are used to identify query candidates. Queries are
then generated by an LLM and progressively filtered through three stages to
retain only context-demanding retrieval items.}
\label{fig:pipeline}
\end{figure*}

The query generation pipeline (Figure~\ref{fig:pipeline}) consists of four main elements: Query Generation, Query Cleaning, Adversarial Filtering and Query Assurance. We go through each in detail below.

\paragraph{Query Generation}\label{sec:query-generator}
To generate queries we use an LLM. Query generation uses GPT-OSS-120B \citet{openai2025gptoss120bgptoss20bmodel}. We chose this model because it was one of the best multilingual open source models. We wanted an open source model to maintain reproducibility. Also, it was one of the cheapest models offered by infrastructure providers. It being cheap is important because we care about creating a scalable pipeline.

The model used medium reasoning-effort and a temperature parameter set to 1. (More on the temperature setting in Appendix~\ref{app:query-mapper-temperature}). The model was set up using vLLM~\citep{kwon2023efficientmemorymanagementlarge} on LUMI (CSC) via DeiC. 

 The model generates queries for query candidates. The model is provided the query candidate (target chunk), its implicit context chunks, and all the context chunks and their implicit context chunks. The model outputs the query and a subset of the context chunks (their internal IDs) that were utilized when crafting the query.
 
 The model is instructed to take the target chunk, specifically consider the content of its context chunks, and craft a query that requires resolving the cross-reference between the two sets to identify the target chunk as the (Figure~\ref{fig:context-chunking}). Thus, this instruction should lead to queries that demand utilizing context outside the target chunk to be solved.
 
 The prompt is presented in the Appendix~\ref{app:query-generator-final-prompt}. The prompt's instructions are placed twice to reinforce instruction-following. This insight follows from our initial experiments with GPT-OSS-20B.

\paragraph{Query Cleaning}
We remove query generations where the generating LLM failed to output a query due to an error e.g. due to timeout. For the Polish dataset, we also remove queries where the target chunk ends with ":" or starts with "-". These chunks did not contain self-contained statements, and thus were problematic to create a query about.

\paragraph{Adversarial Filtering}
We remove all the queries that were either solved by a semantic or keyword retriever. If either retriever ranked the target chunk higher than 10th position, the query was removed. This ensured that the queries that most likely did not require context, were removed.

For the semantic retriever we used BGE-M3 (dense embeddings)~\cite{bge-m3}. It was chosen as it was one of the best multilingual open source retrievers. BGE-M3's max input size was set to 8192 tokens, which fits all chunks in the dataset.

For the keyword retriever we used BM25. It was chosen as BM25 is still a competitive model, while being a different approach from the semantic model. Before passing the text to BM25, we also remove stopwords and perform lemmatization. Stopword removal drops mostly meaningless words from being included in calculations, and lemmatization ensures that the same words with different inflections match each other. To do both we used SpaCy~\cite{honnibal2020spacy} pipelines: pl\_core\_news\_sm for the Polish dataset and da\_core\_news\_sm for the Danish dataset.\

\paragraph{Query Assurance}
As the last step we perform query assurance. The main goal is to further ensure that the query is formulated so that a retriever needs to understand the context chunks. It also checks if the query is solvable at all. The prompt is available in the Appendix~\ref{sec:appendix-query-assurance-final-prompt}. The prompt's instructions are placed twice to reinforce instruction-following.

Query assurance was performed using the same model as in the Query Generation~\ref{sec:query-generator} but with high reasoning.

\textbf{Prompt Tuning}
We tuned the prompt, before we generated the full dataset, to maximise the queries' quality. We performed 8 rounds of manual analysis trying out different prompts. We looked at samples of 30-35 queries, output by the Query Generation. We assessed the precision of the query assurance output. That is, how often we agreed with the assessment of the model. Precision was our focus, to ensure that the queries that actually end up in the dataset require context utilization. For the Polish dataset, during the initial rounds, the precision was about 50-70\%. By the final prompts, the precision had been increased to around 80\%-90\%.

For the Danish dataset, the precision was consistently unsatisfactory. Precision consistently hovered around 50\%, despite repeated attempts to tune the prompt. This was one of the main motivations to include adversarial filtering described earlier. After the filter had been applied, the precision was around 80-90\% as well.

\subsection{Evaluating}
Below we describe the metric used and the two used methods.

\subsubsection{Metrics used}
We report recall@k
\begin{equation}
\begin{aligned}
\text{Recall}@k = \frac{1}{|Q|} \sum_{q \in Q} \frac{|\text{Relevant}_q \cap \text{Retrieved}_q@k|}{|\text{Relevant}_q|} \\
\end{aligned}
\end{equation}
We choose this metric because it is the most interpretable metric that tells how high the target chunk ranks. Providing the recall at multiple $k$ is more informative than the mean rank, because it gives a more granular view of the ranking distribution. It is also unaffected by very low ranked queries.


\subsubsection{Anthropic-style contextual retrieval}\label{sec:anthropic-contextual-retriever}
To solve the task, we implemented Anthropic-style contextual retrieval. In this method, the dataset is pre-processed. An LLM is prompted, with a whole document and a chunk, to describe how the chunk relates to the whole document. Then, the description is prefixed to the chunk's original contents. Finally, a retriever is applied. 

We decided to test this method and use it as a strong baseline, based on the conclusions from \ref{sec:insent-vs-anthropic}.

As the LLM-contextualiser Qwen3-235B-A22B-Instruct-2507-FP8 (non-thinking)~\citep{qwen3technicalreport} was used. It was picked because it was a competitive open source model with a context window of 262k tokens. It could fit our documents as opposed to GPT-OSS-120B, which has a 131k context window. 
The temperature was set to 0 to maximise reproducibility. The contextualising prompt is available in the Appendix~\ref{app:anthropic-prompt}. The prompt's instructions is repeated to reinforce instruction following. 

\subsubsection{Local LLM Contextualisation}\label{sec:local-llm-contextualisation}
We test an approach where multiple contextualisations are created in a sliding window fashion for a chunk. Then, these are all prefixed to the chunk.
The motivation stems from our manual analysis.  Specifically, in both datasets, we found examples of prefixes being paraphrased chunks (or near-copies), or missing the topics the chunk was about. We thought it could be due to an inability of the LLM to locate the chunk in the document, despite the sufficient context window. This phenomenon is known as lost-in-the-middle~\citep{liu2023lostmiddlelanguagemodels}.

In this approach, the document is split into parts, consisting of 32k tokens with 8k token overlap. Then, each part, alongside a chunk, is given to an LLM, prompted to give a local contextualisation of the chunk. We use the same contextualising LLM, and the same prompt as in section~\ref{sec:anthropic-contextual-retriever}. The only difference is one injected line into the prompt: \emph{"Note: you are given a part of the document, not the whole document. Provide context based only on this part."}.

\section{Results}

\subsection{Dataset Throughput}

The pipeline lets only a small fraction of query candidates through, as shown in Table~\ref{tab:dataset-stats}.

\begin{table}[!h]
\centering
\resizebox{\columnwidth}{!}{%
\begin{tabular}{@{}lccc@{}}
\toprule
\textbf{Dataset} & \textbf{Query Candidates} & \textbf{Passed (Count)} & \textbf{Passed (\%)} \\ \midrule
Polish           & 4299                     & 300                     & 7\%                     \\
Danish           & 3111                     & 158                     & 5\%                     \\ \bottomrule
\end{tabular}
}
\caption{Summary of query candidates retained by the final dataset pipeline.}
\label{tab:dataset-stats}
\end{table}

\subsection{Manual analysis of the datasets}\label{sec:manual-analysis-dataset}
To assess the construct validity of the datasets, we randomly
sampled 28 queries from each final dataset and annotated each query along three dimensions: (i) whether selecting the target requires context, (ii) whether the query actually targets the labelled target chunk, and (iii) whether the LLM-selected ``utilised context chunk IDs'' match the chunks truly needed to formulate the query.

(i) and (ii) are not necessarily equivalent. A query might ask for information that is not in the target chunk. In the analysis below, we show examples of queries that ask for information in a context chunk or in chunks that are not present, but would still require context in principle.

Evaluating (iii) is also important because it affects calculations based on the selected context chunks, such as the distance between the target chunk and the context chunk, which we examine in the discussion.

\subsubsection{CRAwLeR-PL dataset}

\paragraph{Context utilisation.} 27 of 28 queries require context. The single failure was due to the target chunk: it contained only explicit references with no substantive content, and the query generator produced a tautology.

\paragraph{Query--target chunk relevance.} 22 of 28 queries both require context and target the labelled target chunk. The six discarded cases consist of the tautology above and five queries following the same pattern. The target chunk specifies \emph{what a Minister shall define}, while the identity of the ``Minister'' appears only in the target chunk's implicit context. That implicit context ends with ``:'', introducing new information that is then specified in the target chunk. An example is shown in Figure~\ref{fig:en-manual-analysis-targeting}. The queries in these cases ask about \emph{the object the Minister should define}, but no chunk contains the answer in the required form. They arguably still target the target chunk semantically, but they fail to follow our definition. The query generation appears to miss the implicit context. In a brief inspection of all 300 queries, we found that when the implicit context ends with ``:'', the generator often, though not always, disregards it and writes a query that does not target the target chunk.
\begin{figure}[!h]
\centering
\begin{tikzpicture}
    \node[draw, rounded corners, fill=gray!5, inner sep=10pt, text width=0.9\linewidth, align=left] {
        \textbf{Query (\texttt{statefireservice\_757}):}\\
        \emph{Who is the competent entity to conduct free medical and psychological examinations of a firefighter delegated to service abroad in a rescue group upon their return?}\\[2ex]
        \textbf{Target chunk:}\\
        \emph{3) the competent entity to conduct the examinations referred to in paragraph 1,}\\[2ex]
        \textbf{Implicit context (containing the noun ``Minister''):}\\
        \emph{Art.~124z. The minister competent for internal affairs shall determine, by means of a regulation: \ldots}
    };
\end{tikzpicture}
\caption{The query asks ``who'' performs the examinations, but it does not target the new information established in the target chunk. A well-formed query should ask about the Minister's obligations, which is to \emph{define} this "who". However, while no chunk contains an answer to this query, the query still requires a context chunk - it contains information from the \emph{ust. 1} context chunk. Original Polish version in Appendix~\ref{app:fig:pl-manual-analysis-targeting}}
\label{fig:en-manual-analysis-targeting}
\end{figure}

\paragraph{Selection of utilised context chunks.} The generator selects \emph{all} truly utilised context chunks in 22 cases; at least one selected chunk is correct in 27 of 28 cases. The six
selection failures all involve queries with multiple truly
utilised context chunks (counts 7, 8, 2, 3, 2, 4). In five of the
six, the unselected chunk is preceded by a chunk ending with
``:''---the same colon-introducer pattern that already requires
filtering during query cleaning (Section~\ref{sec:query-generator}). See example in Figure~\ref{fig:en-manual-analysis-colon-pattern} 
This suggests colon-introducer chunks systematically confuse both
query generation and label attribution. We treat selection errors
as label-quality issues only. Queries with imperfect selections are
not excluded from the 22 retained for downstream analysis, since
the query itself is still well-formed and context-dependent.

\begin{figure}[!h]
\centering
\begin{tikzpicture}
    \node[draw, rounded corners, fill=gray!5, inner sep=10pt, text width=0.95\linewidth, align=left] {
        \textbf{Query (\texttt{obligationtodefend\_2223}):}\\
        \emph{Does the Council of Ministers determine which administrative bodies have the right to impose the obligation to adapt real estate and movable property for the needs of state defense or to perform mobilization tasks for the Armed Forces?}\\[2ex]
        
        \textbf{Target chunk:}\\
        \emph{Art.~222. 1. The Council of Ministers, by means of a regulation, determines the bodies competent to impose the obligations and tasks referred to in Art.~221 sec.~1 and~2.}\\[2ex]
        
        \textbf{Selected context chunk (Colon-introducer pattern):}\\
        \emph{Art.~221. 1. Local government administration bodies, state institutions [...] may be obliged to perform for a fee\textbf{:}}\\[2ex]
        
        \textbf{Missed context chunk (Unselected label):}\\
        \emph{\textbf{[X]} 1) adapting possessed real estate and movable property for the needs of State defense, in a manner that does not change their properties and purpose;}\\[2ex]
        
        \textbf{Selected context chunk:}\\
        \emph{2. Local government administration bodies [...] may be obliged to perform mobilization tasks for the Armed Forces for a fee.}
    };
\end{tikzpicture}
\caption{A partial context selection failure caused by the colon-introducer pattern in CRAwLeR-PL. The generator successfully selects the target chunk and two context chunks, but critically misses the chunk containing bullet point 1. The unselected chunk is immediately preceded by a chunk ending with a colon (``:''), illustrating how this punctuation systematically confuses label attribution when multiple context chunks are required. Original Polish version in Appendix~\ref{app:fig:pl-manual-analysis-colon-pattern}}
\label{fig:en-manual-analysis-colon-pattern}
\end{figure}

\paragraph{79\% pass.} 22 of 28 inspected queries pass both the context-utilisation and target-chunk criteria. The analysis of Anthropic-style contextual retrieval failures is restricted to those queries. An example of a good query is seen in the Figure~\ref{fig:pl-well-formed-query-en}
\begin{figure}[!h]
\centering
\begin{tikzpicture}
    \node[draw, rounded corners, fill=gray!5, inner sep=10pt, text width=0.95\linewidth, align=left] {
        \textbf{Query:}\\
        \emph{On what date is the subsidy for the purchase of textbooks and educational materials for public art schools providing general education transferred?}\\[2ex]
        
        \textbf{Target chunk:}\\
        \emph{2. The specific purpose grant referred to in paragraph 1 shall be provided between 6 May and 15 October.}\\[2ex]
        
        \textbf{Utilized context chunk:}\\
        \emph{Art. 62. 1. To finance the cost of purchasing textbooks, educational materials or exercise materials in the scope referred to in Article 55, paragraph 1, public art schools providing general primary school education run by legal entities that are not local government units and individuals receive, upon request, a targeted subsidy from the state budget. [...]}
    };
\end{tikzpicture}
\caption{An example of a well-formed query targeting a specific statutory condition derived from dispersed contextual information. Original Polish version in Figure~\ref{app:fig:pl-well-formed-query}.}
\label{fig:pl-well-formed-query-en}
\end{figure}

\subsubsection{CRAwLeR-DK dataset}

\paragraph{Context utilisation.} 26 of 28 inspected queries require context to be answered. We note that, in two of these cases, a query targets a context chunk. However, selecting that context chunk still requires awareness of another context chunk.

In two failures, the query also targets a context chunk; however, no other context is necessary to solve it. These two queries come from \emph{erhvervsfondsloven}, one of the hardest documents for us to read. An example is in the Figure~\ref{fig:en-manual-analysis-hijacking}

\begin{figure}[!h]
\centering
\begin{tikzpicture}
    \node[draw, rounded corners, fill=gray!5, inner sep=10pt, text width=0.95\linewidth, align=left] {
        \textbf{Query:}\\
        \emph{Should an interim balance sheet be drawn up for a fund when the joint merger statement has been signed more than six months after the end of the financial year to which the fund's latest annual report relates?}\\[2ex]
        
        \textbf{Target chunk:}\\
        \emph{Chapter 11, Merger of a commercial fund with its wholly owned subsidiary (s), Section on Intermediate Balance Sheet. Section 101. Section 93 applies correspondingly to the fund.}\\[2ex]
        
        \textbf{Utilized context chunk:}\\
        \emph{SECTION 93. If the joint merger statement is signed more than 6 months after the end of the financial year to which the fund's latest annual report relates, an intermediate balance sheet must be drawn up for the fund in question. The fund authority can waive this requirement.}
    };
\end{tikzpicture}
\caption{A query generation failure in CRAwLeR-DK. The query does not target a context chunk, but importantly the context chunk is retrievable on its own, without the use of other context chunks, or the target chunk. The generator failed to include the premise of the target chunk (a merger with a ``wholly owned subsidiary''), leaving the target chunk irrelevant to the generated query. Original Danish version in Appendix~\ref{app:fig:da-manual-analysis-hijacking}}
\label{fig:en-manual-analysis-hijacking}
\end{figure}

\paragraph{Query targeting the target chunk.} 24 of 28 queries target the target chunk. The four ones are the ones already mentioned when discussing the context utilisation. 3 out of 4 failures follow the same ``Under hvilke betingelser \dots'' or ``Hvilke \dots skal \dots'' question pattern: they ask which/who/what object is mentioned in the target chunk, but the object is in a context chunk. An example is in Figure~\ref{fig:en-manual-analysis-cross-ref}.

\begin{figure}[!h]
\centering
\begin{tikzpicture}
    \node[draw, rounded corners, fill=gray!5, inner sep=10pt, text width=0.95\linewidth, align=left] {
        \textbf{Query (\texttt{erhversfondsloven\_\textsection 109\_stk3}):}\\
        \emph{Under what conditions can a commercial foundation that is already in liquidation resume its operations?}\\[2ex]
        
        \textbf{Target chunk:}\\
        \emph{When a foundation has decided to enter liquidation, no decisions may be taken to alter its registered conditions, except for \ldots\ 5) resumption, see \textsection 119, \ldots}\\[2ex]
        
        \textbf{Utilized context chunk:}\\
        \emph{A commercial foundation may be decided resumed if distribution of liquidation proceeds under \textsection 114 has not begun. Resumption is conditional on the appointment of a board and an auditor \ldots, and on a valuation expert's declaration that the foundation's capital is present.}
    };
\end{tikzpicture}
\caption{A query generation failure for CRAwLeR-DK. The target chunk merely cross-references \textsection 119 as one of the changes permitted during liquidation. The substantive conditions live entirely in the context chunk. Original Danish version in the Appendix~\ref{app:fig:da-manual-analysis-cross-ref}.
}
\label{fig:en-manual-analysis-cross-ref}
\end{figure}

\paragraph{Selection of utilised context chunks.} The query generator selects \emph{all} truly utilised context chunks in 24 of 28 cases and at least one in all 28. These failures appear to occur when multiple context chunks are utilised, similarly to the CRAwLeR-PL.

\paragraph{85\% pass.} 24 of 28 inspected queries pass both the context-utilisation and target-chunk criteria. The analysis of Anthropic-style contextual retrieval failures is restricted to those queries. An example of well formed query is in Figure~\ref{fig:da-well-formed-query-en}
\begin{figure}[!h]
\centering
\begin{tikzpicture}
    \node[draw, rounded corners, fill=gray!5, inner sep=10pt, text width=0.95\linewidth, align=left] {
        \textbf{Query (\texttt{barnetslov\_201\_stk1}):}\\
        \emph{Does the state pay the costs of the free counseling and guidance provided to children and young people with reduced physical or mental functioning and their families?}\\[2ex]
        
        \textbf{Target chunk (Pointer):}\\
        \emph{SECTION 201. The state bears the costs of the impartial consulting function according to section 163.}\\[2ex]
        
        \textbf{Utilized context chunk (Payload):}\\
        \emph{SECTION 163. An impartial consultancy scheme provides free advice and guidance in cases of assistance to children and young people with reduced physical or mental functional levels and their families.}
    };
\end{tikzpicture}
\caption{An example of a well-formed query targeting a specific statutory condition derived from dispersed contextual information. Original Danish version in Figure~\ref{app:fig:da-well-formed-query}.}
\label{fig:da-well-formed-query-en}
\end{figure}

\subsection{Anthropic-style Contextual Retrieval}

The results for the CRAwLeR-PL dataset are shown in Table~\ref{tab:polish-results-recall-short} and those for the CRAwLeR-DK dataset are shown in Table~\ref{tab:danish-results-recall-short}.
\begin{table}[!htbp]
\centering
\begin{tabular}{@{}lcc@{}}
\toprule
\textbf{Metric} & \textbf{BM25} & \textbf{BGE-M3} \\ \midrule
Recall@1        & 0.051         & 0.165           \\
Recall@5        & 0.266         & 0.443           \\
Recall@10       & 0.392         & 0.551           \\ \bottomrule
\end{tabular}
\caption{Retrieval performance (Recall@$k$) for CRAwLeR-DK (158 queries). All chunks were augmented following the Anthropic Contextual Retrieval methodology using Qwen3-235B-A22B-Instruct-2507-FP8 (temperature 0).}
\label{tab:danish-results-recall-short}
\end{table}
\begin{table}[!ht]
\centering
\begin{tabular}{@{}lcc@{}}
\toprule
\textbf{Metric} & \textbf{BM25} & \textbf{BGE-M3} \\ \midrule
Recall@1        & 0.067         & 0.193           \\
Recall@5        & 0.377         & 0.490           \\
Recall@10       & 0.470         & 0.587           \\ \bottomrule
\end{tabular}
\caption{Retrieval performance (Recall@$k$) on the CRAwLeR-PL (300 queries). All chunks were augmented following the Anthropic Contextual Retrieval methodology using Qwen3-235B-A22B-Instruct-2507-FP8 (temperature 0).}
\label{tab:polish-results-recall-short}
\end{table}

In addition we inspect whether the contextualisation of the datasets deteriorated the performance on the non-contextual queries, i.e. not let through by the pipeline. The performance does not deteriorate, we notice instead a slight improvement. The results are in the Appendix~\ref{app:anthropic-impact-non-contextual}.

\subsubsection{Manual analysis of Anthropic-style contextual retrieval failures}\label{sec:manual-analysis-anthropic}

\paragraph{CRAwLeR-PL dataset}

We restrict failure-mode analysis to the 22 audited queries that are both well-targeted and context-dependent.

BGE-M3 retrieves the target chunk in the top 10 for 15 of 22 queries; BM25 does so for 14 of 22. 

The failures can be attributed to poor contextualisation. We inspected the four queries missed by
both retrievers. In every case, the Qwen3-generated prefix is
either a near-verbatim paraphrase of the target chunk or otherwise
fails to inject new context (see Figure~\ref{fig:en-manual-analysis-contextualisation}).
\begin{figure}[!h]
\centering
\begin{tikzpicture}
    \node[draw, rounded corners, fill=gray!5, inner sep=10pt, text width=0.95\linewidth, align=left] {
        \textbf{Generated context (\texttt{financingeducation\_688}):}\\[1ex]
        \emph{Art.~66 specifies that in schools run by local government units, the tasks and competencies of the governing body, specified in Art.~63, are performed, respectively, by the commune head (mayor, city president), the county executive board or the regional executive board.}\\[2ex]
        
        \textbf{Target chunk:}\\[1ex]
        \emph{Art.~66. In the case of schools run by local government units, the tasks and competencies of the governing body specified in Art.~63 are performed, respectively, by the commune head (mayor, city president), the county executive board, or the regional executive board.}
    };
\end{tikzpicture}
\caption{Example of a contextualisation failure (near-copy prefix) for CRAwLeR-PL. The generated ``context'' is a paraphrase of the target chunk, and no content is added. Original Polish version in Appendix~\ref{app:fig:pl-manual-analysis-contextualisation}}
\label{fig:en-manual-analysis-contextualisation}
\end{figure}

\begin{figure}[!h]
    \centering
    \includegraphics[width=1\linewidth]{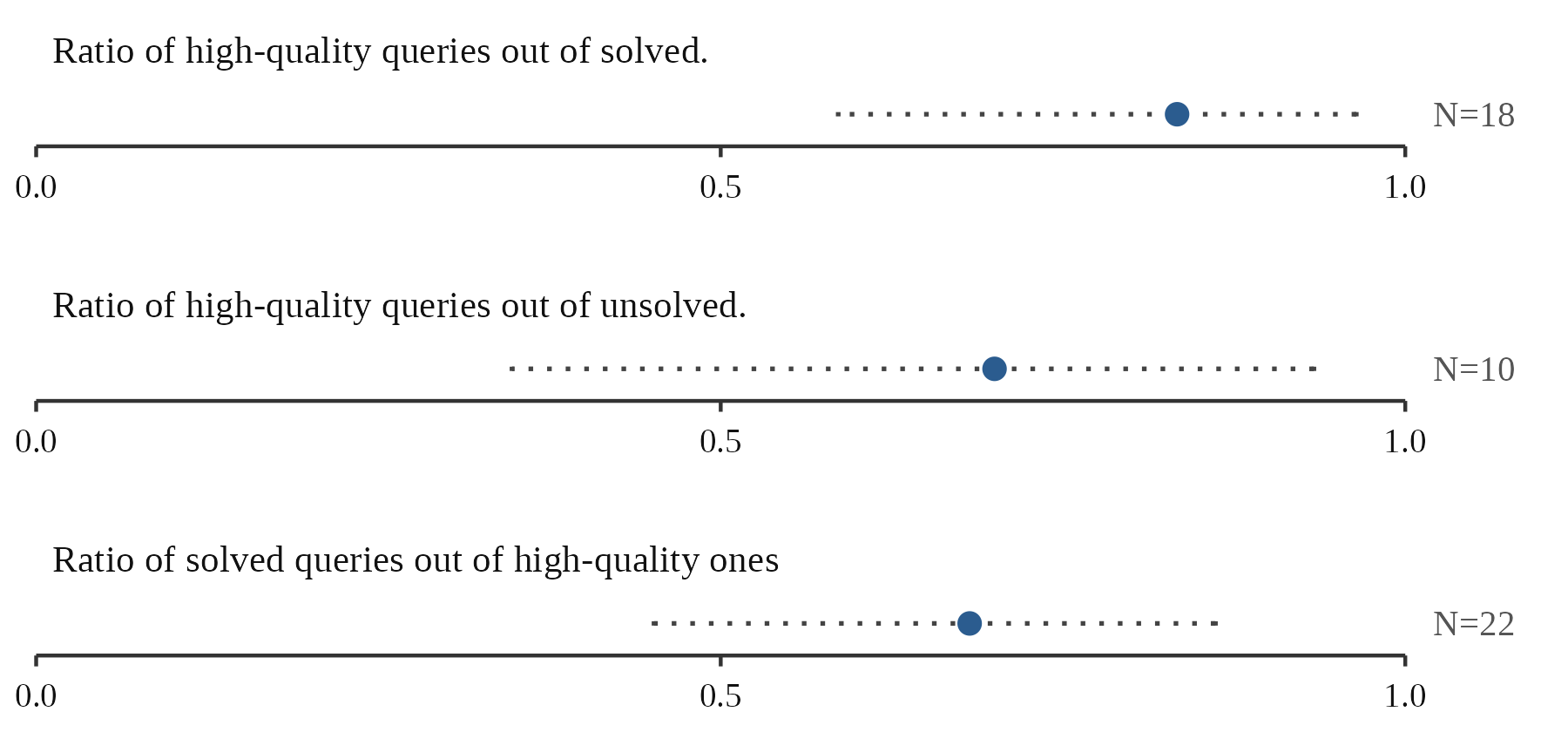}
    \caption{Manual analysis ratios for CRAwLeR-PL with 95\% exact CIs.}
    \label{fig:pl-metrics}
\end{figure}

\paragraph{The unsolved gap contains high-quality queries.} We are also seeing that the baseline is solving mostly high-quality queries. See Figure~\ref{fig:pl-metrics}

\paragraph{Distractor context chunks rank above the target chunk.} Context chunks often rank above the target chunk. At least one context chunk ranks higher than the target in 15 of 22 cases for BM25 and 17 of 22 cases for BGE-M3. All context chunks rank higher than the target in 11 and 10 cases, respectively.

\paragraph{CRAwLeR-DK dataset}
We restrict the failure analysis to the 24 queries that are both well-targeted and context-dependent.

BGE-M3 retrieves the target chunk in the top 10 for 13 of 24 queries; BM25 does so for 13 of 24. 

We inspected the 8 queries missed by both retrievers. 
In 3 of 8, the contextualisation describes the wrong topic, often reusing a generic prefix shared across sibling chunks (Figure~\ref{fig:en-manual-analysis-wrong-prefix}). In 5 of 8, the prefix is under-specified and does not describe the important context sufficiently. In a few cases, for example, the prefix contains the cross-reference of a relevant context chunk rather than describing the chunk's contents (Figure~\ref{fig:en-manual-analysis-underspecific-prefix}).
\begin{figure}[!h]
\centering
\begin{tikzpicture}
    \node[draw, rounded corners, fill=gray!5, inner sep=10pt, text width=0.95\linewidth, align=left] {
        \textbf{Query (\texttt{serviceloven\_\textsection 82b\_stk2}):}\\
        \emph{Can a municipal board decision to grant a time-limited subsidy for individual help, care, support or training -- which may be given for up to 6 months and only on the assessment of improvement or prevention of deterioration -- be appealed to another administrative authority?}\\[2ex]
        
        \textbf{Target chunk:}\\
        \emph{The municipal board's decision under stk.~1 cannot be appealed to another administrative authority.}\\[2ex]
        
        \textbf{Generated context (Wrong-topic prefix):}\\
        \emph{This section concerns municipal board decisions on general offers with an activating and preventive aim, including the setting of guidelines for target groups and payment for the offers.}
    };
\end{tikzpicture}
\caption{A contextualisation failure (wrong-topic prefix) for CRAwLeR-DK. The generated prefix incorrectly describes the section as concerning ``general offers'' instead of ``time-limited individual'' support. Original Danish version in Appendix~\ref{app:fig:da-manual-analysis-wrong-prefix} }
\label{fig:en-manual-analysis-wrong-prefix}
\end{figure}
\begin{figure}[!h]
\centering
\begin{tikzpicture}
    \node[draw, rounded corners, fill=gray!5, inner sep=10pt, text width=0.95\linewidth, align=left] {
        \textbf{Query (\texttt{almenboligloven\_\textsection 58b\_stk1}):}\\
        \emph{Can the municipal board allocate a vacant elderly home or nursing-home place when it has entered into an agreement that the public housing organisation takes over the allocation of the public elderly homes?}\\[2ex]
        
        \textbf{Target chunk:}\\
        \emph{\textsection~58~b. Elderly and persons with disabilities who need an elderly home, a nursing-home place or protected housing are placed on a waiting list in the municipality of residence \ldots\ Vacant homes are allocated by the municipal board, \textbf{except per \textsection~55, stk.~1}, \ldots\ to the persons with the greatest need for the home in question and then to those who have been on the waiting list the longest.}\\[2ex]
        
        \textbf{Generated context (under-specific prefix):}\\
        \emph{This section is part of chapter 4 on letting and allocation of public housing, and specifically the subsection on free choice of elderly homes. The section regulates placement on waiting lists for elderly and persons with disabilities seeking an elderly home, a nursing-home place or protected housing, and the order of allocation of vacant homes based on need and waiting time.}
    };
\end{tikzpicture}
\caption{A contextualisation failure (underspecific prefix) in CRAwLeR-DK. The prefix correctly describes the chunk's topic (waiting-list rules and allocation order) but does not surface the crucial exception clause \emph{except per \textsection~55, stk.~1} inside the chunk. The query targets exactly that exception. The labelled context chunk \textsection~55, stk.~1 contains the agreement scenario described in the query almost verbatim, rendering it a very good distractor example. Unlike wrong-topic prefix failures, the prefix here is factually correct, but it omits the crucial detail. Original Danish version in Appendix~\ref{fig:dk-manual-analysis-underspecific-prefix}.}
\label{fig:en-manual-analysis-underspecific-prefix}
\end{figure}

\begin{figure}[!h]
    \centering
    \includegraphics[width=1\linewidth]{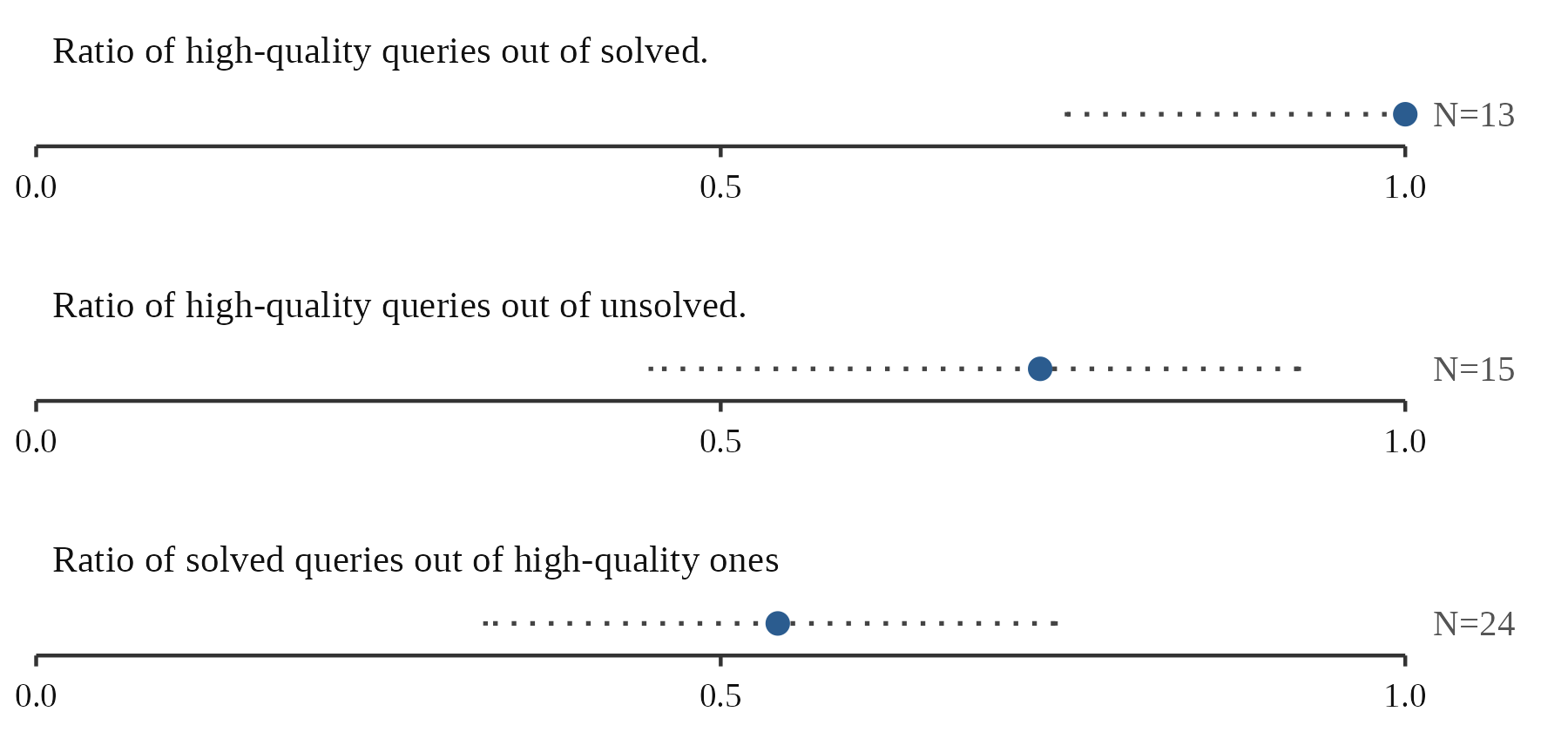}
    \caption{Manual analysis ratio for CRAwLeR-DK with 95\% exact CIs.}
    \label{fig:dk-metrics}
\end{figure}

\paragraph{The unsolved gap contains high-quality queries.} We are also seeing that the baseline is solving mostly high-quality queries. See Figure~\ref{fig:dk-metrics}

\paragraph{Distractor context chunks rank above the target chunk.}
The context chunks frequently outrank the target chunk. At least one context chunk ranks higher than the target in 21 of 24 cases for BM25 and 16 of 24 cases for BGE-M3. All context chunks rank higher than the target in 16 and 14 cases, respectively.

\subsection{Local Contextualisation}
We run an experiment comparing local contextualisation (defined in Section~\ref{sec:local-llm-contextualisation}) with global contextualisation (Anthropic-style contextual retrieval). 

We do not observe substantial differences, especially given the small number of queries. Table~\ref{tab:waterfall-recall-bge-stacked}.
\begin{table}[!h]
\centering
\resizebox{\columnwidth}{!}{%
\begin{tabular}{@{}lccccc@{}}
\toprule
\textbf{Method} & \textbf{R@1} & \textbf{R@5} & \textbf{R@10} & \textbf{R@50} & \textbf{R@100} \\ \midrule
Global & 0.138 & 0.552 & 0.621 & 0.793 & 0.897 \\
Local  & 0.172 & 0.414 & 0.586 & 0.897 & 0.966 \\ \bottomrule
\end{tabular}
}
\caption{Comparison of recall metrics between global and local contextualisation using the BGE-M3 embedder on a CRAwLeR-PL document \texttt{obligationtodefend}, with 29 queries.}
\label{tab:waterfall-recall-bge-stacked}
\end{table}

\subsubsection{Distance results}\label{sec:distance-results}
\begin{figure}[!h]
    \centering
    \includegraphics[width=0.48\textwidth]{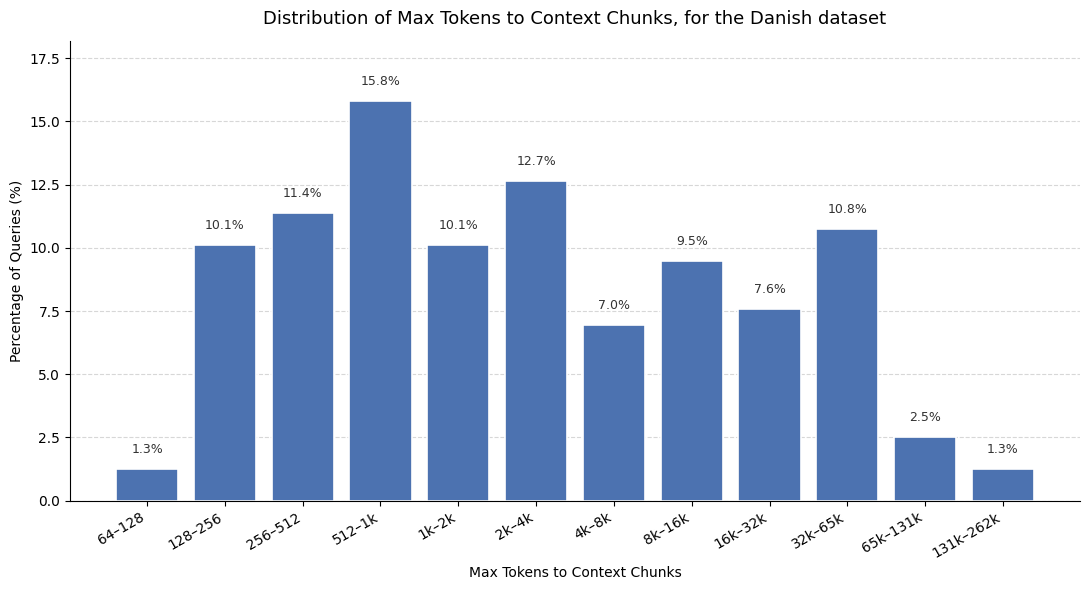}
    \caption{Token distances from the target chunk to the farthest labelled context chunk in the CRAwLeR-DK dataset.}
    \label{fig:token-distances-danish}
\end{figure}

\begin{figure}[!h]
    \centering
    \includegraphics[width=0.48\textwidth]{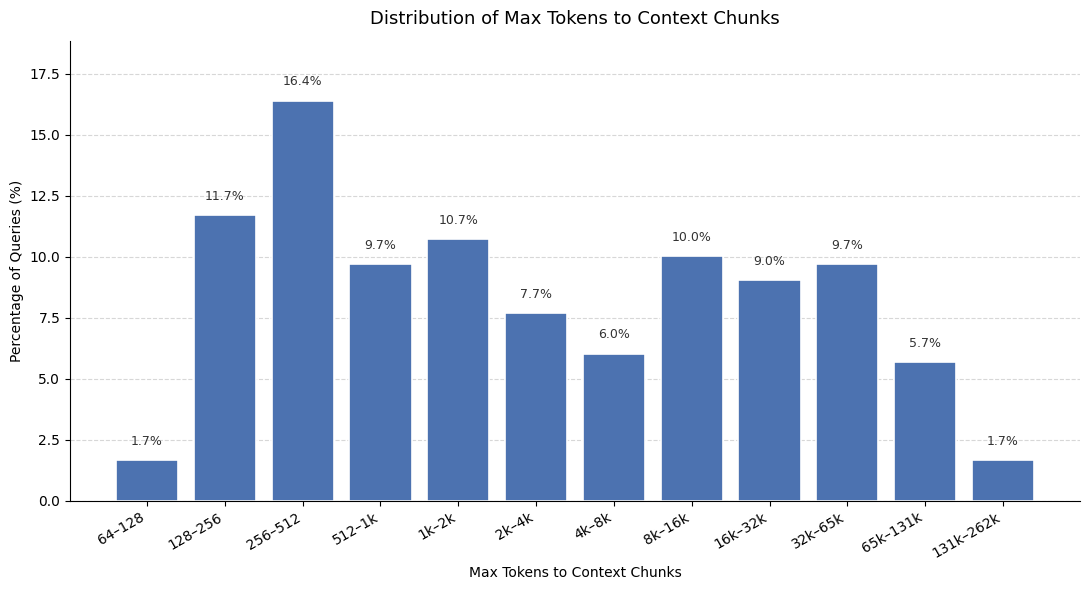}
    \caption{Token distances from the target chunk to the farthest labelled context chunk in the CRAwLeR-PL dataset.}
    \label{fig:token-distances-polish}
\end{figure}

We compute the distribution of token distances to the farthest context chunk selected by the query generator for the CRAwLeR datasets. See Figures~\ref{fig:token-distances-danish} and~\ref{fig:token-distances-polish}. We discuss the implications in the discussion.

\section{Ablation}
We perform an ablation of the best model. We obtain a 2x2 table, where we switch the contextualising LLM and the semantic embedder to weaker versions. We picked multilingual-e5-small (mE5\textsubscript{small})~\citet{wang2024multilinguale5textembeddings}, because it is expected to perform worse than BGE-M3 (dense). Evaluated on 16 languages in MIRACL multilingual retrieval benchmark it obtains average nDCG@10 of 60.8~\citet{wang2024multilinguale5textembeddings} against M3's 69.2~\citet{bge-m3}. Its large version scores average nDCG@10 of 34.2 against M3's 52.5 multilingual long-doc retrieval on the MLDR test set~\citet{bge-m3}.

To obtain a comparison independent from the context window, we compare both models limited to 512 tokens input. 

As the weaker contextualising LLM, we picked Qwen3-30B-A3B-Instruct-2507-FP8 as it was a less complex model from the same family as the initial contextualising LLM.
\begin{table}[!h]
\centering

\label{tab:ablation-study}
\begin{tabular}{@{}llccc@{}}
\toprule
\textbf{Context.} & \textbf{Retriever} & \textbf{R@1} & \textbf{R@5} & \textbf{R@10} \\ \midrule
Weak & Weak & 0.130 & 0.307 & 0.420 \\
Strong & Weak & 0.200 & 0.480 & 0.590 \\
Weak & Strong & 0.120 & 0.303 & 0.417 \\
Strong & Strong & 0.193 & 0.490 & 0.587 \\ \bottomrule
\end{tabular}
\caption{2x2 ablation study on the CRAwLeR-PL dataset comparing two different contextualisers: Qwen3-235B-A22B-Instruct-2507-FP8 (Strong) and Qwen3-30B-A3B-Instruct-2507 (Weak) at temperature=0, and two different embedders: BGE-M3 (Strong) and $mE5_{small}$ (Weak). Both are limited to 512 tokens. R is recall}
\end{table}

\begin{table}[!h]
\centering
\label{tab:ablation-study-danish}
\begin{tabular}{@{}llccc@{}}
\toprule
\textbf{Context.} & \textbf{Retriever} & \textbf{R@1} & \textbf{R@5} & \textbf{R@10} \\ \midrule
Weak & Weak & 0.101 & 0.304 & 0.411 \\
Strong & Weak & 0.184 & 0.468 & 0.551 \\
Weak & Strong & 0.082 & 0.285 & 0.399 \\
Strong & Strong & 0.165 & 0.443 & 0.551 \\ \bottomrule
\end{tabular}
\caption{2x2 ablation study evaluated on the CRAwLeR-DK dataset, comparing two different contextualisers: Qwen3-235B-A22B-Instruct-2507-FP8 (Strong) and Qwen3-30B-A3B-Instruct-2507 (Weak) at temperature=0, and two different embedders: BGE-M3 (Strong) and $mE5_{small}$ (Weak). Both are limited to 512 tokens.}
\end{table}

In the Appendix~\ref{app:ablation-study} there are tables comparing BGE-M3 with input length set to its maximum of 8192 tokens. The difference is negligible confirming that the comparison for 512 tokens measures the model's understanding, not an advantage stemming merely from 8k tokens input size of BGE-M3. Another reason why the comparison was not hindered by the different context windows, was that the most of the CRAwLeR chunks were less than 512 tokens long (Appendix ~\ref{app:data-sources}).

Surprisingly the weak retriever is sometimes even slightly better than BGE-M3. These results suggest that a contextualising LLM is substantially more important for the retrieval than a retriever.

\section{Discussion}

\subsection{Are the CRAwLeR queries good quality?}

We discuss query quality along the three dimensions described in the manual analysis.

The queries do appear to require context: 27/28 for CRAwLeR-PL. The single CRAwLeR-PL tautology is labelled as a failure, but it appears to be caused by one problematic chunk rather than by a systematic issue. For CRAwLeR-DK, the corresponding figure is 26/28. There are two more examples where the queries do require context. However, they are not directed at the target chunk, meaning they fail to function as a task item for CRAwLeR. 

Once we also consider the requirement that queries should target the specified target chunk, we end up with around 80\% of high-quality queries for both datasets. Is it sufficient? The question is rather, what change we can observe given the noise. One could be certain that any improvement of 20\% or more, included solving at least one high-quality query. Other approach could be probabilistic. If we knew the ratio of high-quality queries per 1 solved queries, and assumed is constant, then we could estimate the expectation of solved high-quality queries. For the baseline such a ratio seems to be positive. However, validity of this approach would benefit from more analysis.

Furthermore, both datasets contain systematic errors that could be addressed fast. In the CRAwLeR-PL dataset, the Query Generation struggles with using the implicit context of the target chunk. We suspect it may be due to the query generation prompt not sufficiently emphasizing the potential importance of the implicit context. A simple solution would be including an example of how to construct a valid query that effectively uses implicit context. Another solution, without regenerating queries, would be to remove all queries whose implicit context ends with '':'', effectively removing the problematic queries mentioned in Section~\ref{sec:manual-analysis-dataset}. In the CRAwLeR-DK dataset, all four problematic queries target a context chunk. In this case, a quick round of relabelling could be applied to the target chunks; alternatively, all queries starting with phrases such as ``Under which conditions'' (who/what/which queries) could simply be removed, as these often target context chunks.

The targeted context chunk selection is usually correct. While it is imperfect, we note that this is a secondary annotation and does not affect query quality with respect to the CRAwLeR task. Moreover, the insights about the distributions of token distances to the context chunks (Section ~\ref{sec:distance-results}) should not be substantially affected: we have no evidence of a systematic relationship between erroneously selected context chunks and their token distance from the target chunk.

\subsection{Are the datasets solved?}
The manual analysis shows promising results. For Polish dataset, the unsolved gap composes in 70\% of the high-quality queries. Also, out of all high quality queries 32\% were unsolved. It means, there is a room to measure an improvement. For Danish dataset, the conclusions are similar, giving 73\% and 48\% accordingly. 
It suggests that datasets would be suitable to measure a non-trivial improvement. However, we need to note that the results, are based on limited samples sizes (Figures~\ref{fig:pl-metrics} and~\ref{fig:dk-metrics}).

Based on the inspected contextualisations, we conclude that the contextualising LLM is a major bottleneck in both datasets and that the current R@10 numbers indeed reflect the method limitations. The gap to solving these benchmarks is substantial, and it is attributable to contextualisation quality, suggesting opportunities for better methods.


In addition, our manual analysis showed that context chunks usually still rank higher then the context chunks, even when the target chunk appears in the top ten. This suggests that the retrievers still confuses the target. \\ 
The frequency of this occurring for the whole dataset could be computed in future work.

The local contextualisation did not help, butit still could be a viable approach. Manual analysis of the errors was is to be done and could reveal a simple fix.

\subsection{Ablation study}
Our ablation showed that the contextualising LLM is important for the quality of the retrieval, and that the retriever is substantially less important. This is a positive results in terms of implying that the datasets require effective context utilization to be solved.

However, we note a limitation. The comparable performance of the "Weaker" to the "Stronger" model could be due to the fact that mE5\textsubscript{small} could be comparable with BGE-M3 in Polish and Danish, despite being worse on average across different languages. We note we found that mE5's \emph{large} version has a comparable R@100 in Polish and Danish on MKQA dataset~\citet{bge-m3}. However, we still think that in the expectation the mE5\textsubscript{small} was a worse model, justifying the choice.

\subsection{Scalability of contextual pipeline}
One thing that motivated this project was a specific gap in the existing literature: there is no efficient, automated method for generating queries that require non-trivial context utilization. With our pipeline in hand, it is worth considering what we have actually achieved in this regard, and what the implications are for scalability. Our pipeline is strict by design. It applies both an adversarial filter and an assurance prompt. The query generation prompt itself also enforces a strict set of requirements on how each query must be constructed (see appendix). Table \ref{tab:dataset-stats} shows the resulting throughput. For the Polish dataset, we ran the pipeline on 4299 query candidates, of which 300 (7\%) passed. For the Danish dataset, we ran it on 3111 query candidates, of which 158 (5\%) passed. A first implication is for cost. For our specific setup, the cost of producing a final dataset is roughly the cost of running the pipeline on a single item, multiplied by the desired number of queries, multiplied by 20 given the approximately 5\% throughput. A second implication is more positive. We think our pipeline does a better job of enforcing context use than the synthetic pipeline used in ConTeb, discussed earlier in the literature review. Our qualitative analysis of model predictions also suggests that an encouragingly high proportion of surviving queries do require context to be solved.

There are some interesting design trade-offs worth considering. The first concerns how strongly one values context utilization being required at all. In a setting where it is less critical, our pipeline is likely overkill given the cost. But our pipeline is more applicable when the goal is to be confident the phenomenon is being measured, with scores that can be interpreted as performance on context utilization across legal cross-references. A second, more complicated trade-off concerns the choice of model used for query generation. Counterintuitively, it may be cheaper to use a more expensive model. If the model follows instructions more accurately and produces queries less likely to fail, throughput goes up. The total cost of generating a dataset may then drop despite the higher per-call cost. This cost-throughput tradeoff has recently been formalized by \citet{fu2025costeffective}, which treats synthetic-data generation as selecting among models with different costs and post-filter rewards, optimizing “utility = reward/cost.” In this view, the relevant quantity is not cost per generated item, but cost per retained valid item. 

The same logic is worth considering for the assurance model, although the direction of effect is less clear. A more capable assurance model might fail more queries, because it catches more subtle errors or technicalities a weaker model would miss. Or it might pass more queries, if some current failures are caused by the assurance model incorrectly rejecting a valid query. From our several rounds of prompt tuning, where we manually examined the output of the pipeline, it seems the former effect would dominate.

\subsection{Discussing Construct Validity}
\subsubsection{Content Validity}
As noted in the literature review, previous benchmarks for the broader task of context-aware chunk retrieval have not addressed construct validity in much detail. A recent survey on construct validity in language model benchmarks found that, among papers claiming to measure long-context capabilities, fewer than 20\% substantially discuss the construct validity of their benchmark \citep{constructvaliditymetasurvey}. Our work claims to measure context utilization, or more specifically, the ability of models to conduct cross-reference-aware context utilization for chunk retrieval in legal documents. This phenomenon of interest was operationalized through the CRAwLeR task. This section discusses whether our benchmark measures that capability, and to what extent. In doing so, we aim to improve on the limited construct-validity discussion in the related benchmarks discussed in the literature review. For now, we  focus primarily on content validity, often considered a feature of construct validity. That is, how well our task items in CRAwLeR-DK and CRAwLeR-PL actually represent the theoretical task space of our phenomenon of interest \citep{contentvalidity}.

To be clear, the task space is the set of all possible task items that could plausibly fall under the phenomenon of interest \citep{constructvaliditymetasurvey}. According to the work by \citet{constructvaliditymetasurvey}, proper coverage of this space is central to content validity, and construct validity in practice. We identify three main limitations here. First, our pipeline covers only some ways in which cross-reference dependencies can be used to create task items. Second, our dataset contains a small number of documents, which limits representativeness even within our specific domain: Danish and Polish laws. Third, the dataset does not cover other domains where similar cross-references occur. 

We believe legal cross-reference dependencies can support several types of  task items, relevant for our phenomenon, but our pipeline and CRAwLeR targets only some of them. The survey by \citet{constructvaliditymetasurvey} recommends deliberate sampling methods, such as targeted or strategic sampling, to cover the task space more adequately. It also notes that this applies not only to sampling from existing datasets, but also to sampling from the theoretical set of possible task items. Our pipeline treats the referencing chunk as the target and the referenced chunk or chunks as the context. However, the direction could also be reversed: a referenced chunk may be the target, while a later referencing chunk provides the relevant context. The assumption would be that the referencing chunk says something about the referenced chunk. This would allow it to play the role of a context chunk, rather than a target chunk. 

Another under-represented case is queries that depend on multiple referenced chunks. Our query generation prompt did not require the model to target a single referenced chunk, but more than 80\% of generated queries did so. More targeted sampling could have produced better coverage of such queries. 

A third dimension is context chain length. We limited cross-reference chains to a single hop because models did not reliably use longer chains in our experiments, and the added chunks often introduced noise. This made the examples harder to interpret and analyse. Deeper cross-reference chains are still a natural extension if query generation can be made more reliable. Other relevant dimensions include chunk size and the distance between target and context chunks. The distances between target and context chunks can be seen in figure \ref{fig:token-distances-danish}. It's unclear whether this distribution is an accurate estimation of the true distribution in our domain. The answer is probably not, given our small number of documents. However, what we can tell, is that our task items mostly cover distances between target chunks and context chunks uniformly. The exception is more extreme distances (65k tokens or beyond). This should be considered when interpreting results on CRAwLeR-Dk or CRAwLeR-PL.

Another content-validity limitation of our work is related to domain representativeness. Our dataset contains relatively few documents. Prior work has noted that a recurring challenge in long-context utilization datasets is finding concentrated sources of sufficiently long documents \citep{difficultypaper}, and we encountered this directly. If the domain is defined narrowly as Danish and Polish laws, the small number of documents still limits representativeness within that domain. This creates a standard content-validity concern: if the sampled items do not adequately span the phenomenon, benchmark scores cannot be cleanly interpreted as performance on that phenomenon. The third limitation is related. Similar cross-references occur in other domains, but our dataset does not cover them. Because other domains may differ in structure, language, and retrieval difficulty, they plausibly contain important task items that our benchmark misses. However, if they do not qualify as part of the legal domain, they arguably do not limit our content validity. This is because our phenomenon of interest, cross-reference-aware context utilization for chunk retrieval in legal documents, is definitionally only concerned with the legal domain.

A further construct-validity concern is confounding. Confounders affect what is actually measured by changing model failure modes. They may cause a model to fail for reasons unrelated to the intended phenomenon, or allow it to succeed without using the intended capability. Following the recommendation of \citet{constructvaliditymetasurvey}, we also analysed model failures to see whether they aligned with the phenomenon or were driven by confounders. For each reviewed query, we inspected the top 10 retrieved chunks. We looked for false negatives and lexical overlap. We consider false negatives as chunks that a model could reasonably treat as relevant but that were not labelled as correct. Lexical overlap means high surface-level word overlap that could let a model succeed without using the cross-reference.
Across 56 reviewed queries, 28 per dataset, we found no false negatives. This likely reflects both our prompt tuning and properties of the legal domain. Because we used an adversarial filter, and because we wanted to avoid creating false negatives, we did not instruct the query-generation model to use vague terms merely to reduce lexical overlap. The legal domain also uses specific and exact language, which likely helped reduce false negatives. At the same time, other contextual clues may still help models identify the correct chunk, such as coreference, multi-hop context, or broader structural cues. It is difficult to separate these effects fully. As a result, our dataset may partly measure other forms of context utilization in addition to cross-reference resolution. Additionally, no problematic patterns of lexical overlap were found.

\subsubsection{Other Features of Construct Validity}
Content validity has been the most useful lens for this work. It clarifies what the benchmark measures, what it leaves out, and what future work should address. It also gives the clearest basis for interpreting what the results cover. Still, other facets of construct validity are worth considering briefly. Face validity, meaning whether the test appears to represent the intended phenomenon at face value \citep{facevalidity}, is reasonably strong. CRAwLeR appears to represent the intended phenomenon because the definitions of target chunks, context chunks, and query instructions all support this: a query is designed so that the target chunk can only be identified by using information from a cross-referenced context chunk. Ecological validity, meaning how well the test reflects real-world settings \citep{ecological}, is more limited. The datasets were primarily designed to enforce context use, not to closely mimic how legal experts search legal corpora. The task resembles ad hoc retrieval, but the generated queries may not match expert search behavior on corpora like ours. For example, the query generation prompt asks the model to avoid specific section numbers, to avoid giving hints, but a legal expert might want to include them. We did not think that convergent validity, meaning whether results align with other tests that claim to measure a similar phenomenon \citep{convergent}, could be reasonably assessed. As established, we consider the other benchmarks for context-aware chunk retrieval to have construct validity issues, meaning comparisons would be less meaningful and interesting. Other facets of construct validity, such as divergent validity \citep{convergent}, are also relevant, but would require larger methodological undertakings. So, ultimately we leave those unaddressed.

\subsection{Long-Context Utilization}
\subsubsection{Trivial Perspective}
Beyond context utilization, we claim our dataset measures long-context utilization for our phenomenon. Recent work has called for more datasets that demand genuinely difficult long-context behaviour \citep{difficultypaper}, which makes it worth examining whether our work meets these expectations. As far as we know, there is no widely agreed-upon definition of long-context utilization in the information retrieval setting we are working in. Two benchmarks aimed at dense embedders, LongEmbed \citep{zhu-etal-2024-longembed} and LoCoV1 \citep{10.5555/3692070.3693819}, treat the term as referring to tasks that demand long-context windows mostly by virtue of input size. Under that reading, our dataset qualifies trivially: document lengths exceed 32k tokens, which is beyond the input limits of most current state-of-the-art dense embedders.

\subsubsection{Scope and Dispersion}
A more interesting lens is one often invoked in discussions of retrieval and long-context capabilities of generative language models. Previous work by \citet{difficultypaper} proposes that, for a task to genuinely qualify as long-context, it should be high along at least one of two orthogonal axes: dispersion or scope. Briefly, dispersion captures how hard it is to find the necessary information in the context, quite literally how spread out that information is across the document, while scope captures how much necessary information there is to find. 

For our task, only the target chunk has to be retrieved at the end of the day, so the question of scope reduces to how much information is needed to find that single chunk. The answer is not much. As discussed above, most queries target a single chunk, and individual chunks are relatively small in tokens compared to the document as a whole. Scope, then, is generally low in our datasets. 

Dispersion is more interesting. The paper outlines three sub-aspects under which dispersion can be high: sparsity (the relevant information is interspersed with non-required information, hiding in a crowd), obscurity (the relevant information is obscured behind contextual dependencies that need to be resolved), and a lack of redundancy (the same information is not restated in multiple places). On the first, our task generally scores high: the chunks relevant to any given query are a small portion of the document. On the second, our task also scores high: contextual dependencies must be resolved to recover the meaning of a target chunk, and none of the chunks containing the necessary information are individually sufficient. We note, however, that cross-references are a relatively explicit and detectable form of context dependency compared to other types, since they can in principle be picked out with regexes or similar surface tools, which limits the obscurity claim somewhat. On the third, our manual analysis did not find redundant restatements of the relevant information, and we expect this not to apply in general, since legal language tends to be highly specific. 

By the input-size perspective of long-context utilization, our datasets qualify trivially. By the dispersion/scope perspective, they qualify on the dispersion axis but not the scope axis. We think the latter is a more meaningful way to look at whether our datasets require long-context utilization.

\subsection{Lacking legal expertise}

This project depended heavily on the analysis and interpretation of legal texts, and we have no legal expertise. This type of missing application-domain expertise is a known source of downstream data-quality failures in AI pipelines \citep{lackoflegalexpertise1}. We expect this to have affected our results in several ways. Most importantly, it bears on our qualitative manual analysis, and the percentage of queries we determined as actually demanding context. Our reading of each query in relation to its target and context chunks may have been wrong in some cases, which puts those numbers into question. This is particularly relevant because NLP dataset quality management directly affects whether resulting models and evaluations are reliable \citep{lackoflegalexpertise2}. Likewise it would negatively affect our work's construct validity. The second concerns prompt tuning. 

We conducted several rounds of tuning based on the problems we observed in the pipeline outputs. Legal experts would likely have written better prompts. The benefits could include more intelligent query design and broader coverage of the task space. They could also have produced queries that reflect a practical legal setting more accurately, improving ecological validity. Failure rates at quality assurance might also have dropped, lowering pipeline cost by raising throughput beyond what we report in table \ref{tab:dataset-stats}. The third concerns the correctness of our parsing code. As prior work has pointed out, official rules for how laws must be drafted offer limited guidance on how cross-references should be written, and therefore detected \citep{sannier2017automated}. Our source corpora were no exception. We therefore relied on our own reading of the legal texts when constructing the parsers for chunking and reference extraction. Because of our lack of expertise, the parsers may systematically miss certain cross-reference patterns. This would worsen our content validity.

\section{Future Work}
\subsection{Achieving better coverage of the task space}
We will now bring up some ways that could improve coverage of the task space for cross-reference-aware context utilization for chunk retrieval in legal documents. Achieving these would further improve content validity.

\paragraph{Context chains.} One promising approach is making proper use of cross-reference chains. In the legal domain we investigated, a single chunk will often reference a second chunk, which in turn references a third, and so on. Such chains can be used to construct queries that demand multi-hop reasoning, where intermediate conclusions serve as a foundation for further reasoning. We did not manage to establish this in the current dataset. The current chain length is capped at 1, when providing context chunks to the query-generation model. An early version of the pipeline did not limit the chain length. This resulted in the query-generation model receiving a large, noisy amount of context at query generation time. The significant increase in input would also make costs of our pipeline rise. Also, our manual review showed the model rarely used these longer chains when writing queries, and the noisier context made manual review itself harder. Doing it properly will likely require more deliberate prompting that forces the model to reason over more than one hop. We expect this to come with trade-offs. More queries will likely fail quality assessment, since such complex queries are harder to generate. The model given the assurance prompt will also be put under more pressure, and may require a stronger underlying model. If it can be done, and the model accurately reports which context chunks it targeted, we expect that chain length becomes both a useful dimension of difficulty and a useful attribute for analysis. One could, for instance, examine whether performance drops as the number of required hops increases. 

\paragraph{Inter-document cross-references.} A second direction is expanding the dataset to cover inter-document cross-references. In the legal domain we investigated, a fraction of cross-references point to other documents rather than to other parts of the same document. Incorporating these would add both difficulty and broader coverage of the task space. Doing so requires more intelligent detection of which document a reference points to. It also requires that the referenced document and its chunks are actually part of the dataset. This would also stress methods that rely on within-document information for chunk contextualization, such as late chunking and Anthropic's contextualization approach, since the relevant context may sit outside the target chunk’s document. Beyond inter-document references, we would also encourage expanding the task space along several axes mentioned in the discussion: more queries that target multiple context chunks, greater variation in the distance between target and context chunks, broader coverage of the domain by including more documents, and extension of the task framework into other domains where other cross-references occur.

\paragraph{Semantic intent.} A third direction is incorporating classification of the semantic intent behind each cross-reference. Previous work has already explored such classification with promising results \citep{semanticintentclassification}. An example of how this would help is that while tuning the pipeline prompts, we manually reviewed many generated queries. We found that some legal cross-references did not lend themselves to a context-dependent query under our current prompt. We expect that tailoring the query generation prompt to the semantic intent of the reference could improve generation quality. This would, in turn, reduce cost by lowering the rate of items filtered out at quality assurance. One concrete example is cross-references whose semantic intent is to introduce an exception to a rule defined in the referencing chunk. The model often misunderstood these and produced queries that were conceptually flawed. As for other parts of the project, legal expertise would be helpful here. For instance, by facilitating identification of the relevant intent categories and writing prompts suited to each. 

\subsection{Encouraging narrower scope}
We encourage future work on chunk retrieval to avoid treating context utilization as a single, broad capability, at least for now. Instead, benchmarks and datasets should target narrower and better-defined sub-components of the phenomenon. Context utilization can involve several different abilities, such as resolving references, following discourse structure, using coreference, or combining information across chunks. Measuring all of these at once makes benchmark construction difficult and creates risks for construct validity, especially when task items are repurposed from datasets that were not designed for this purpose. A narrower scope reduces generalizability, but it makes the resulting scores easier to interpret. At the current stage, where benchmarks for context-dependent chunk retrieval remain limited, we believe this trade-off is worthwhile. More focused datasets can help clarify the task space, support more precise definitions, and make it easier to understand which forms of context use current models can and cannot handle. Once several such datasets exist, future benchmarks may be better positioned to measure the broader phenomenon by combining or repurposing task items across sub-components.

\section{Conclusion}

We asked two questions in the introduction: if one can build
a benchmark whose scores can be interpreted as evidence of
cross-reference-aware context utilization, and whether current
contextualisation methods solve such a benchmark. The first
question receives a positive answer for the narrow phenomenon we
operationalize. The second receives a negative one.

CRAwLeR-DK and CRAwLeR-PL hold up under manual analysis. Approximately
80\% of randomly sampled queries indeed target the labelled
target chunk and require context to be answered, and the remaining
failures form named, systematic patterns, that are addressable through prompt-level fixes during query generation or, in the CRAwLeR-DK case, relabelling. To
our knowledge, this are the first datasets in context-aware chunk
retrieval to provide construct-validity evidence at this level of
granularity, which makes the datasets' scores interpretable as
evidence about the intended phenomenon rather than only about
the dataset.

The datasets are not solved. Best Recall@10 reaches
55\% on CRAwLeR-DK and 59\% on CRAwLeR-PL with BGE-M3 under
Anthropic-style contextualization, meaningfully above chance, but still having a space to measure the improvements. Crucially, this gap is set by the contextualising
LLM, not by the retriever. Our ablation shows that swapping the
contextualiser for a weaker model is the main reason for the
performance loss, while swapping the dense embedder for a smaller
one is costless. Failure analysis on queries missed by both
retrievers confirm that. The generated prefix is
either factually wrong about the chunk's topic or correct but
omits the chunk's important detail. Even when the target chunk
is retrieved in the top ten, the context chunks routinely
rank higher than it. That is, the contextualiser is not making the target chunk distinguishable from its own references.

The limit of the tested methods is set
by what a contextualising LLM surfaces about each chunk, not by
the retriever's discriminative ability. Progress on this task
should focus there. For new benchmark design, we think
that narrow, manually-analysed scope produces more interpretable
scores than broad coverage with weak validation, at least at the
current stage where benchmarks for context-aware chunk
retrieval are still maturing.

\section{Acknowledgments}
This work was carried out as the authors' bachelor's thesis at the IT University of Copenhagen. We would like to thank Professor Ratish Puduppully for his academic supervision.

\clearpage
\bibliography{custom}

\appendix
\section*{Appendix}
\section{Temperature in Query Generation and Query Assurance}\label{app:query-mapper-temperature}
We acknowledge that the temperature should have been set up to 0 to ensure more reproducible results. (It was supposed to be so, but due to the oversight default parameter was used.). However, we note the impact is not so large. Firstly, the queries were saved, and any results on them can be recomputed. Secondly, the prompt was already highly structured constraining the model decisions.

\section{Prompts}
\subsection{Query Generation Final Prompt}\label{app:query-generator-final-prompt}
\begingroup
\scriptsize
\itshape

\textless prompt\textgreater

Your task is to generate a query for a retrieval dataset focused on contextual dependencies, i.e., cases where the target chunk to be selected as positive, understanding of the context is required.

You are given a \textbf{target chunk} from a legal text and the \textbf{context chunks} it references. You may also be given \textbf{Target Implicit Context Chunks}.

\textbf{Definitions:}
\begin{itemize}
  \setlength{\itemsep}{0pt}
  \setlength{\parskip}{0pt}
  \setlength{\parsep}{0pt}
\item \textbf{Citation, explicit reference} - pinpoint citation
\item \textbf{Target chunk}: A chunk that contains at least one pinpoint citation to other chunk(s) within the same document.
\item \textbf{Context chunks}: The chunks explicitly referenced by the target chunk. They also contain their own implicit context chunks, at the bottom, which may be helpful to understand the context chunks.
\item \textbf{Target Implicit Context Chunks}: chunks adjacent to the target chunk (i.e., the immediately preceding and maybe also following chunks in the document, coming from the same legal paragraph), if they exist. They may or may not be explicitly referenced by the target chunk. 
\item \textbf{targets the target chunk} - it means it asks about a statement i.e. a new piece of information that is established in the target chunk. It does not mean simply referencing the target chunk.
\end{itemize}

\vspace{0.5em}\hrule\vspace{0.5em}

\textbf{Your task}: Generate exactly one realistic, natural-language retrieval query satisfying ALL of the following conditions:

Key conditions:
\begin{enumerate}
  \setlength{\itemsep}{0pt}
  \setlength{\parskip}{0pt}
  \setlength{\parsep}{0pt}
\item \textbf{Targets the target chunk} — The answer to the query is a piece of information established in the target chunk, that is dependent on some context chunks.  
    \begin{itemize}
      \setlength{\itemsep}{0pt}
      \setlength{\parskip}{0pt}
      \setlength{\parsep}{0pt}
    \item The query should not target any statement that is made in the context chunks or neighboring chunks.  
    \end{itemize}
\item \textbf{However, the retriever requires context chunks} - for the target chunk to be selected as positive, at least one of the context chunks is required. The target chunk will \textit{not} be retrieved based on its own content alone.
    \begin{itemize}
      \setlength{\itemsep}{0pt}
      \setlength{\parskip}{0pt}
      \setlength{\parsep}{0pt}
    \item This means that the query should contain information that is only found in the context chunks, and that is required to understand the target chunk.
    \end{itemize}
\end{enumerate}

Clarifications:
\begin{enumerate}
  \setcounter{enumi}{2}
  \setlength{\itemsep}{0pt}
  \setlength{\parskip}{0pt}
  \setlength{\parsep}{0pt}
\item \textbf{Query does not need to be dependent on all context chunks} - it is fine to use at least one context chunk (and obviously a target chunk), but it is not necessary to use all of them. 

\item \textbf{The query must not contain explicit references to the context chunks} - the point is that if you are crafting a query, that contains the relationship between the target chunk and context chunks, refer to the substance of the context chunk, not simply place its identifier like "Article 5 paragraph 2 point 3 letter a". This totally misses the point.

\item \textbf{No multiple questions hidden in the query}: create a query poiting to a SINGLE statement made in the target chunk.

\item \textbf{Use of the Target Implicit Context Chunks}: They might be helpful to understand the target chunk.

\item \textbf{Avoid unnecessary lexical overlap}: The query should not simply repeat the same words as the target chunk, but rather use different wording to express the same meaning. This is to ensure that the retriever is not just matching keywords, but actually understanding the content. However, ensure that the words that are crucial, especially domain-specific terms, are not changed to the point of losing the interpretability and the clear connection between the query and the target chunk.

\item \textbf{Multiple statement in the target chunk} - If the target chunk contains multiple statements, the query should be about only one such statement, which is dependent on some context chunks.

\item \textbf{Pinpoint citations may be abbreviated}: A target chunk may cite a context chunk in full (e.g., "art. 45 ust. 2 pkt 1 ustawy") or in abbreviated form when the context is nearby (e.g., "ust. 2 pkt 1", or even just "pkt 1"). Context chunks are themselves prefixed with their own (sometimes shorter) identifier. All context chunks provided have been explicitly referenced by the target chunk — your job is to match each citation to the right chunk, not to judge whether the reference exists.
\end{enumerate}

\vspace{0.5em}\hrule\vspace{0.5em}

\textbf{Examples}:

1. In Danish:

Target chunk: \S~10. Stk. 2. For ansatte omfattet af \S~2, nr. 3, finder stk. 1, nr. 2, ikke anvendelse.

Context chunks: \S~2. Ydelser efter denne bekendtgørelse tilkommer ansatte, der er udsendt til tjeneste uden for landet:
3) for at sikre driften af en offentlig institution i dens tjenesteområde.
\S~10. Stk. 1. Særtillægget udgør:
4) 300 kr. pr. døgn under tjeneste i områder med særlige belastninger.
No neighboring chunks.

Good query: Får ansatte, der er udsendt til tjeneste uden for landet for at sikre driften af en offentlig institution i dens tjenesteområde, særtillægget på 300 kr. pr. døgn under tjeneste i områder med særlige belastninger?

Bad query: Hvilke ansatte får ikke særtillægget på 300 kr. pr. døgn under tjeneste i områder med særlige belastninger?

\vspace{0.5em}
2. In Polish:

Target chunk: 2. Do żołnierzy, o których mowa w \S~2 pkt 2 lit. e, nie stosuje się ust. 1 pkt 2.

Context chunks: \S~2. Należności pieniężne określone przepisami niniejszego rozporządzenia przyznaje się żołnierzom zawodowym: \\ 2) skierowanym do pełnienia zawodowej służby wojskowej poza granicami państwa: \\ e) w celu zabezpieczenia funkcjonowania jednostki wojskowej użytej zgodnie z przepisami ustawy z dnia 17 grudnia 1998 r. o zasadach użycia lub pobytu Sił Zbrojnych Rzeczypospolitej Polskiej poza granicami państwa, w rejonie jej działania albo zapewnienia organizacji, funkcjonowania i sprawowania działalności kontrolnej tej jednostki wojskowej w rejonie jej działania. \\ \S~10. 1. Stawka dodatku wojennego, o którym mowa w art. 468 ust. 5 ustawy, wynosi: \\ 2) od 0,03 do 0,05 najniższego uposażenia - za każdy dzień wykonywania zadań w strefie działań wojennych w warunkach związanych z bezpośrednim udziałem w akcjach o charakterze bojowym, akcjach zapobiegania aktom terroryzmu lub ich skutkom albo pełnieniem służby patrolowej, ochronnej lub z udziałem w konwojach.

No neighboring chunks.

Good query: Czy żołnierze zawodowi, którzy są skierowani do pełnienia służby wojskowej poza granicami państwa w celu zabezpieczenia funkcjonowania jednostki wojskowej, otrzymują dodatek wojenny za każdy dzień wykonywania zadań?

Bad query: Jacy żołnierze nie otrzymują dodatku wojennego za każdy dzień wykonywania zadań?

In both cases, the good query targets a statement in the target chunk, which is dependent on the context chunks. The bad query, on the other hand, does not target any statement in the target chunk.

Good queries also contain much of the content from the context chunks, so that a simple lexical retriever would get confused.

\vspace{0.5em}\hrule\vspace{0.5em}

\textbf{Tactic}:

Read the target chunk and list the statements it makes. Identify which statements depend on context chunks via pinpoint citations. If several qualify, pick one — the query will target only that statement. Read neighboring chunks if needed to disambiguate the target chunk.

Resolve the citations. For each pinpoint citation in your chosen statement, match it to the corresponding context chunk. Citations may be abbreviated (e.g., "ust. 2 pkt 1" referring back to an article established earlier), and context chunks carry their own identifier prefix that may be shorter still — align them by working from the most specific component outward. Once matched, mentally substitute the citation with the substantive content of the context chunk. Use the context chunks' own neighboring chunks (provided beneath them, with internal IDs to help you map them) only to interpret the context — never as a target.

Draft the query so that:
\begin{itemize}
  \setlength{\itemsep}{0pt}
  \setlength{\parskip}{0pt}
  \setlength{\parsep}{0pt}
\item it asks about the chosen statement in the target chunk,
\item it carries enough substance from the context chunk(s) that retrieval cannot succeed on the target chunk alone,
\item it does not name or cite the context chunks by identifier,
\item it phrases things differently from the target chunk's wording where possible, while keeping domain-specific terms intact,
\item it asks a single question, ideally opening with Do/Does/Is/Are/Under which conditions/When (avoid Who/What/Which, which tend to ask for entities defined in the context chunks rather than rules established in the target chunk).
\end{itemize}

\vspace{0.5em}\hrule\vspace{0.5em}

\textbf{Output:}

Return ONLY a valid JSON object with exactly two keys. Do not include markdown formatting outside the JSON, no explanation, no preamble, and no quotation marks.

\begin{verbatim}
{
  "utilized_context_chunk_ids": ["ID_1", "ID_2"],
  "query": "a single question in the same language as the input legal text"
}
\end{verbatim}

\vspace{0.5em}\hrule\vspace{0.5em}

\textbf{Input:}
\begin{itemize}
  \setlength{\itemsep}{0pt}
  \setlength{\parskip}{0pt}
  \setlength{\parsep}{0pt}
\item \textbf{Target chunk}: \textless target\_chunk\textgreater\ \{\{chunk\}\} \textless/target\_chunk\textgreater
\item \textbf{Target Implicit Context Chunks}: \textless target\_implicit\_context\_chunks\textgreater\ \{\{impl\_context\_chunks\}\} \textless/target\_implicit\_context\_chunks\textgreater
\item \textbf{Context chunks}: \textless context\_chunks\textgreater\ \{\{context\_chunks\}\} \textless/context\_chunks\textgreater
\end{itemize}

\textless /prompt\textgreater

\endgroup

Then, the prompt is repeated  (without the inputs).

\subsection{Query Assurance Final Prompt}\label{sec:appendix-query-assurance-final-prompt}
\begingroup
\scriptsize
\itshape

\textless prompt\textgreater

The task is to evaluate the quality of a retrieval dataset focused on contextual dependencies i.e. for the target chunk to be selected as positive, understanding of the context is required.

You are given a \textbf{query}, \textbf{context chunks} and a \textbf{target chunk} that references these context chunks, and also \textbf{Target Implicit Context Chunks} (optionally).

\textbf{Definitions:}
\begin{itemize}
  \setlength{\itemsep}{0pt}
  \setlength{\parskip}{0pt}
  \setlength{\parsep}{0pt}
\item \textbf{Query} - a question about the target chunk. However, for the target chunk to be selected as positive, understanding some of the context chunks and/or Target Implicit Context Chunks is required. This is what you are going to evaluate.
\item \textbf{Target chunk}: The chunk that contain pinpoint citations referring to other chunks.
\item \textbf{Context chunks}: The chunk(s) cited by the target chunk. At the bottom, these might contain their own implicit context chunks to help you better understand the context chunks.
\item \textbf{Target Implicit Context Chunks}: Chunk(s) that might help you understand the target chunk. E.g. adjacent to the target chunk (i.e., the immediately preceding and maybe also following chunks in the document, coming from the same legal paragraph), or section/chapter titles; if they exist. They may or may not be cited by the target chunk. They may or may not be required to understand the target chunk, but they should not be the basis for the query unless they are clearly required to understand the target chunk (e.g. the target chunk is mentioned as the 'previous' or 'next' chunk or e.g. the Target Implicit Context Chunks are one sentence, hierarchically split into clauses by order, and the target chunk cannot be understood without preceding clauses i.e. chunks).
\item \textbf{Citation, explicit reference} - pinpoint citation
\item \textbf{Perfect retriever and perfect reader}: assume an ideal retriever and an ideal reader; if all necessary information is present in the provided chunks, they will identify the target chunk as positive and interpret it correctly.
\item \textbf{Sufficient}: a chunk or set of chunks is sufficient if a perfect retriever, together with a perfect reader, could identify the target chunk as positive given the query using only those chunks and no additional information. If the target chunk might directly answer the query, but the provided chunk set does not contain the information needed to interpret it as an answer, then that chunk set is not sufficient. 'About the statement' means the query asks about a piece of information established in the target chunk, not merely about locating or naming the target chunk.
\item \textbf{(Likely required) Context chunks IDS} - these are the (internal) IDs of the context chunks that \textit{are likely} required to be seen to correctly select the target chunk as positive given the query. They are provided, as not all the context chunks might be required to solve this particular query. They will help you to double check if the context was required or not. Since the labelling is not 100\% perfect, you should also always other context chunks, but this is the first place to look to check. This is just a hint.
\end{itemize}

\vspace{0.5em}\hrule\vspace{0.5em}

\textbf{Your task}: Evaluate the query against the five criteria below. Reason briefly through each before giving a verdict.

\textbf{Criterion 1 — Query-target chunk relevance}\\
Is the query about the information stated in the target chunk, rather than about information stated only in the context chunks or Target Implicit Context Chunks?\\
The query may mention conditions or categories supplied by context, but the thing being asked about must be the information established in the target chunk.\\
Consider the context and the Target Implicit Context Chunks to assess this criterion.\\
$\rightarrow$ Answer: [Yes / No]

\vspace{0.5em}
\textbf{Criterion 2 — Target chunk alone retrievability}\\
Is the target chunk on its own, without using the context chunks or Target Implicit Context Chunks, sufficient to be identified as positive by a perfect retriever given the query?\\
$\rightarrow$ Answer: [Yes / No]

\vspace{0.5em}
\textbf{Criterion 3 — All combined}\\
Are the target chunk together with the context chunks, and the Target Implicit Context Chunks, sufficient to identify the target chunk as positive by a perfect retriever given the query?\\
$\rightarrow$ Answer: [Yes / No]

\vspace{0.5em}
\textbf{Criterion 4 — Contextual retrieval advantage}\\
Would this item reward a retrieval model that contextualizes the target chunk with the relevant context chunks, rather than a non-contextual retriever that could retrieve the target chunk without context mainly through lexical overlap, semantic similarity to the target chunk alone, entity or name overlap, number or date overlap, or other distinctive surface-form matching?\\
$\rightarrow$ Answer: [Yes / No]

\vspace{0.5em}
\textbf{Final verdict: [Yes / No]}\\
Yes = Criterion 1 is Yes, Criterion 2 is No, Criterion 3 is Yes, Criterion 4 is No, Criterion 5 is Yes\\
No = Otherwise

\vspace{0.5em}\hrule\vspace{0.5em}

\textbf{Output format:}

Return a JSON object with the following keys:
\begin{itemize}
  \setlength{\itemsep}{0pt}
  \setlength{\parskip}{0pt}
  \setlength{\parsep}{0pt}
\item "criterion\_1" through "criterion\_4": one sentence summary per criterion
\item "answer\_to\_query": brief phrase or sentence answering the query
\item "verdict": exactly "Yes" or "No"
\end{itemize}

\vspace{0.5em}\hrule\vspace{0.5em}

\textbf{Examples}:

1. In Danish:

Target chunk: \S~10. Stk. 2. For ansatte omfattet af \S~2, nr. 3, finder stk. 1, nr. 2, ikke anvendelse.

Context chunks: \S~2. Ydelser efter denne bekendtgørelse tilkommer ansatte, der er udsendt til tjeneste uden for landet:
3) for at sikre driften af en offentlig institution i dens tjenesteområde.
\S~10. Stk. 1. Særtillægget udgør:
4) 300 kr. pr. døgn under tjeneste i områder med særlige belastninger.
No neighboring chunks.

Good query: Får ansatte, der er udsendt til tjeneste uden for landet for at sikre driften af en offentlig institution i dens tjenesteområde, særtillægget på 300 kr. pr. døgn under tjeneste i områder med særlige belastninger?

Bad query: Hvilke ansatte får ikke særtillægget på 300 kr. pr. døgn under tjeneste i områder med særlige belastninger?

\vspace{0.5em}
2. In Polish:

Target chunk: 2. Do żołnierzy, o których mowa w \S~2 pkt 2 lit. e, nie stosuje się ust. 1 pkt 2.

Context chunks: \S~2. Należności pieniężne określone przepisami niniejszego rozporządzenia przyznaje się żołnierzom zawodowym: \\ 2) skierowanym do pełnienia zawodowej służby wojskowej poza granicami państwa: \\ e) w celu zabezpieczenia funkcjonowania jednostki wojskowej użytej zgodnie z przepisami ustawy z dnia 17 grudnia 1998 r. o zasadach użycia lub pobytu Sił Zbrojnych Rzeczypospolitej Polskiej poza granicami państwa, w rejonie jej działania albo zapewnienia organizacji, funkcjonowania i sprawowania działalności kontrolnej tej jednostki wojskowej w rejonie jej działania. \\ \S~10. 1. Stawka dodatku wojennego, o którym mowa w art. 468 ust. 5 ustawy, wynosi: \\ 2) od 0,03 do 0,05 najniższego uposażenia - za każdy dzień wykonywania zadań w strefie działań wojennych w warunkach związanych z bezpośrednim udziałem w akcjach o charakterze bojowym, akcjach zapobiegania aktom terroryzmu lub ich skutkom albo pełnieniem służby patrolowej, ochronnej lub z udziałem w konwojach.

No neighboring chunks.

Good query: Czy żołnierze zawodowi, którzy są skierowani do pełnienia służby wojskowej poza granicami państwa w celu zabezpieczenia funkcjonowania jednostki wojskowej, otrzymują dodatek wojenny za każdy dzień wykonywania zadań?

Bad query: Jacy żołnierze nie otrzymują dodatku wojennego za każdy dzień wykonywania zadań?

In both cases, the good query targets a statement in the target chunk, which is dependent on the context chunks. The bad query, on the other hand, does not target any statement in the target chunk.

Good queries also contain much of the content from the context chunks, so that a simple lexical retriever would get confused.

\vspace{0.5em}\hrule\vspace{0.5em}

\textbf{Input:}
\begin{itemize}
  \setlength{\itemsep}{0pt}
  \setlength{\parskip}{0pt}
  \setlength{\parsep}{0pt}
\item Query: \textless query\textgreater \{\{query\}\} \textless /query\textgreater
\item Target chunk: \textless target\_chunk\textgreater \{\{chunk\}\} \textless /target\_chunk\textgreater
\item Target Implicit Context Chunks: \textless target\_implicit\_context\_chunks\textgreater \{\{impl\_context\_chunks\}\} \textless /target\_implicit\_context\_chunks\textgreater
\item Context chunks: \textless context\_chunks\textgreater \{\{context\_chunks\}\} \textless /context\_chunks\textgreater
\item Context chunk IDs: \textless context\_chunk\_ids\textgreater \{\{context\_chunk\_ids\}\} \textless /context\_chunk\_ids\textgreater
\end{itemize}

\textless /prompt\textgreater
\endgroup

Then, the prompt is repeated (without the inputs).

\subsection{Anthropic Contextual Retrieval Prompt}\label{app:anthropic-prompt}
\begingroup
\scriptsize
\itshape
System: You are a precise context augmenter 

User:

\textbf{Goal:} Give a context to situate this chunk in the context of the document for the purposes of improving search retrieval of the chunk.

\textbf{Instructions:} 
\begin{itemize}[nosep]
\item Please give a succinct context to situate this chunk within the overall document for the purposes of improving search retrieval of the chunk. 
\item Answer only with the context and nothing else.
\end{itemize}

\textbf{Context:} 

\textless document\textgreater \{\{document\}\}\textless /document\textgreater

Here is the chunk we want to situate within the whole document:

\textless chunk\textgreater \{\{chunk\}\}\textless /chunk\textgreater

\textbf{Instructions reminder:} \emph{(Instructions list is placed here again)} \\
Answer in the language of the document. (Document and the chunk are in the same language).

\endgroup
\subsection{Sliding Window Aggregate Prompt}\label{app:sliding-window-consolidation-prompt}
\begingroup
\scriptsize
\itshape

System: You are a precise context augmenter

User:

\textbf{Goal:} You are given several context snippets that were each written to situate the same chunk within different parts of a long document. Merge them into one succinct context that captures all distinct information, removes duplication, and is suitable for improving search retrieval.

\textbf{Instructions:}
\begin{itemize}[nosep]
\item Answer only with the merged context and nothing else.
\end{itemize}

\textless contexts\textgreater \{\{numbered\}\} \textless /contexts\textgreater

\textless chunk\textgreater \{\{chunk\}\} \textless /chunk\textgreater

Answer in the language of the chunk.

\endgroup

As this was significantly shorter, the instructions were not repeated.

\section{Ablation study}\label{app:ablation-study}
\begin{table}[h]
\centering

\label{tab:ablation-study-polish-8192}
\begin{tabular}{@{}llccc@{}}
\toprule
\textbf{LLM} & \textbf{Retrieval} & \textbf{R@1} & \textbf{R@5} & \textbf{R@10} \\ \midrule
Weak & Weak & 0.130 & 0.307 & 0.420 \\
Strong & Weak & 0.200 & 0.480 & 0.590 \\
Weak & Strong & 0.123 & 0.303 & 0.420 \\
Strong & Strong & 0.193 & 0.490 & 0.587 \\ \bottomrule
\end{tabular}
\caption{2x2 ablation study evaluated on the Polish dataset, comparing two different contextualisers: Qwen3-235B-A22B-Instruct-2507-FP8 (Strong) and Qwen3-30B-A3B-Instruct-2507 (Weak) at temperature=0, and two different embedders: BGE-M3 (Strong) and $mE5_{small}$ (Weak). Note that BGE-M3 used its max input length of 8192 tokens, whereas $mE5_{small}$ was limited to 512 tokens.}
\end{table}

\begin{table}[h]
\centering
\label{tab:ablation-study-danish-8192}
\begin{tabular}{@{}llccc@{}}
\toprule
\textbf{Augmentation} & \textbf{Retrieval} & \textbf{R@1} & \textbf{R@5} & \textbf{R@10} \\ \midrule
Weak & Weak & 0.101 & 0.304 & 0.411 \\
Strong & Weak & 0.184 & 0.468 & 0.551 \\
Weak & Strong & 0.082 & 0.285 & 0.399 \\
Strong & Strong & 0.165 & 0.443 & 0.551 \\ \bottomrule
\end{tabular}
\caption{2x2 ablation study evaluated on the Danish dataset, comparing two different contextualisers: Qwen3-235B-A22B-Instruct-2507-FP8 (Strong) and Qwen3-30B-A3B-Instruct-2507 (Weak) at temperature=0, and two different embedders: BGE-M3 (Strong) and $mE5_{small}$ (Weak). Note that BGE-M3 used its max input length of 8192 tokens (no cap), whereas $mE5_{small}$ was limited to 512 tokens.}
\end{table}
For the Danish dataset tiny differences are seen from top 50 or top 100 metrics only.

\FloatBarrier

\section{Manual analysis examples}\label{app:manual-analysis-examples}
\begin{figure}[!h]
\centering
\begin{tikzpicture}
    \node[draw, rounded corners, fill=gray!5, inner sep=10pt, text width=0.9\linewidth, align=left] {
        \textbf{Query (\texttt{statefireservice\_757}):}\\
        \emph{Kto jest podmiotem właściwym do przeprowadzania bezpłatnych badań lekarskich i psychologicznych strażaka delegowanego do służby poza granicą państwa w grupie ratowniczej po jego powrocie?}\\[2ex]
        \textbf{Target chunk:}\\
        \emph{3) podmiot właściwy do przeprowadzania badań, o których mowa w ust.~1,}\\[2ex]
        \textbf{Implicit context (containing the noun ``Minister''):}\\
        \emph{Art.~124z. Minister właściwy do spraw wewnętrznych określi, w drodze rozporządzenia\textbf{:} \ldots}
    };
\end{tikzpicture}
\caption{Original Polish version of Figure~\ref{fig:en-manual-analysis-targeting}}
\label{app:fig:pl-manual-analysis-targeting}
\end{figure}
\begin{figure}[!h]
\centering
\begin{tikzpicture}
    \node[draw, rounded corners, fill=gray!5, inner sep=10pt, text width=0.95\linewidth, align=left] {
        \textbf{Query (\texttt{obligationtodefend\_2223}):}\\
        \emph{Czy Rada Ministrów określa, które organy administracji mają prawo nakładać obowiązek dostosowania nieruchomości i rzeczy ruchomych do potrzeb obrony państwa lub wykonywania zadań mobilizacyjnych na rzecz Sił Zbrojnych?}\\[2ex]
        
        \textbf{Target chunk:}\\
        \emph{Art.~222. 1. Rada Ministrów, w drodze rozporządzenia, określa organy właściwe do nakładania obowiązków i zadań, o których mowa w art.~221 ust.~1 i~2.}\\[2ex]
        
        \textbf{Selected context chunk (Colon-introducer pattern):}\\
        \emph{Art.~221. 1. Terenowe organy administracji rządowej, instytucje państwowe [...] mogą być zobowiązane do odpłatnego\textbf{:}}\\[2ex]
        
        \textbf{Missed context chunk (Unselected label):}\\
        \emph{\textbf{[X]} 1) dostosowania posiadanych nieruchomości i rzeczy ruchomych do potrzeb obrony Państwa, w sposób niezmieniający ich właściwości i przeznaczenia;}\\[2ex]
        
        \textbf{Selected context chunk:}\\
        \emph{2. Terenowe organy administracji rządowej [...] mogą być zobowiązane do odpłatnego wykonania zadań mobilizacyjnych na rzecz Sił Zbrojnych.}
    };
\end{tikzpicture}
\caption{Original Polish version of Figure~\ref{fig:en-manual-analysis-colon-pattern}}
\label{app:fig:pl-manual-analysis-colon-pattern}
\end{figure}
\begin{figure}[!h]
\centering
\begin{tikzpicture}
    \node[draw, rounded corners, fill=gray!5, inner sep=10pt, text width=0.95\linewidth, align=left] {
        \textbf{Generated context (\texttt{financingeducation\_688}):}\\[1ex]
        \emph{Art.~66 określa, że w szkołach prowadzonych przez jednostki samorządu terytorialnego zadania i kompetencje organu prowadzącego, określone w art.~63, wykonują odpowiednio wójt (burmistrz, prezydent miasta), zarząd powiatu lub zarząd województwa.}\\[2ex]
        
        \textbf{Target chunk:}\\[1ex]
        \emph{Art.~66. W przypadku szkół prowadzonych przez jednostki samorządu terytorialnego zadania i kompetencje organu prowadzącego określone w art.~63 wykonuje odpowiednio wójt (burmistrz, prezydent miasta), zarząd powiatu, zarząd województwa.}
    };
\end{tikzpicture}
\caption{Original Polish version of Figure~\ref{fig:en-manual-analysis-contextualisation}}
\label{app:fig:pl-manual-analysis-contextualisation}
\end{figure}
\begin{figure}[!h]
\centering
\begin{tikzpicture}
    \node[draw, rounded corners, fill=gray!5, inner sep=10pt, text width=0.95\linewidth, align=left] {
        \textbf{Query (\texttt{financingeducation\_653}):}\\
        \emph{W jakim terminie przekazywana jest dotacja przeznaczona na zakup podręczników i materiałów edukacyjnych dla publicznych szkół artystycznych realizujących kształcenie ogólne?}\\[2ex]
        
        \textbf{Target chunk:}\\
        \emph{2. Dotacja celowa, o której mowa w ust. 1, jest przekazywana w terminie od dnia 6 maja do dnia 15 października.}\\[2ex]
        
        \textbf{Utilized context chunk:}\\
        \emph{Art. 62. 1. Na sfinansowanie kosztu zakupu podręczników, materiałów edukacyjnych lub materiałów ćwiczeniowych w zakresie, o którym mowa w art. 55 ust. 1, publiczne szkoły artystyczne realizujące kształcenie ogólne w zakresie szkoły podstawowej prowadzone przez osoby prawne niebędące jednostkami samorządu terytorialnego oraz osoby fizyczne otrzymują, na wniosek, dotację celową z budżetu państwa. [...]}
    };
\end{tikzpicture}
\caption{Original Polish version of Figure~\ref{fig:pl-well-formed-query-en}.}
\label{app:fig:pl-well-formed-query}
\end{figure}
\begin{figure}[!h]
\centering
\begin{tikzpicture}
    \node[draw, rounded corners, fill=gray!5, inner sep=10pt, text width=0.95\linewidth, align=left] {
        \textbf{Query (\texttt{erhversfondsloven\_§101\_stk1}):}\\
        \emph{Skal der udarbejdes en mellembalance for en fond, når den fælles fusionsredegørelse er underskrevet mere end seks måneder efter udløbet af det regnskabsår, som fondens seneste årsrapport vedrører?}\\[2ex]
        
        \textbf{Target chunk:}\\
        \emph{Kapitel 11, Fusion af en erhvervsdrivende fond med dens helejede datterselskab(er), afsnit om Mellembalance. §~101. §~93 finder tilsvarende anvendelse på fonden.}\\[2ex]
        
        \textbf{Utilized context chunks:}\\
        \emph{§~93. Hvis den fælles fusionsredegørelse er underskrevet mere end 6 måneder efter udløbet af det regnskabsår, som fondens seneste årsrapport vedrører, skal der for den pågældende fond udarbejdes en mellembalance. Fondsmyndigheden kan dispensere fra dette krav.}\\[1ex]
        \emph{Stk.~2. Mellembalancen, der skal udarbejdes i overensstemmelse med årsregnskabsloven, må ikke have en opgørelsesdato, der ligger mere end 3 måneder forud for underskrivelsen af den fælles fusionsredegørelse. Mellembalancen skal være revideret.}
    };
\end{tikzpicture}
\caption{Original Danish version of Figure~\ref{app:fig:da-manual-analysis-hijacking}}
\label{app:fig:da-manual-analysis-hijacking}
\end{figure}
\begin{figure}[!h]
\centering
\begin{tikzpicture}
    \node[draw, rounded corners, fill=gray!5, inner sep=10pt, text width=0.95\linewidth, align=left] {
        \textbf{Query (\texttt{erhversfondsloven\_\textsection 109\_stk3}):}\\
        \emph{Under hvilke betingelser kan en erhvervsdrivende fond, der allerede er i likvidation, genoptage sin virksomhed?}\\[2ex]
        
        \textbf{Target chunk:}\\
        \emph{Stk. 3. Når der i en fond er truffet beslutning om at træde i likvidation, kan der ikke træffes beslutning om ændring af de registrerede forhold vedrørende fonden bortset fra \ldots\ 5) genoptagelse, jf.~\textsection 119, \ldots}\\[2ex]
        
        \textbf{Utilized context chunk:}\\
        \emph{\textsection 119. En erhvervsdrivende fond kan besluttes genoptaget, hvis uddeling af likvidationsprovenu efter \textsection 114 ikke er påbegyndt. Det er en betingelse for genoptagelsen, at der udpeges en bestyrelse og en revisor \ldots, og at der udarbejdes en erklæring af en vurderingsmand \ldots, om, at fondens grundkapital er til stede. \ldots}
    };
\end{tikzpicture}
\caption{Original Danish version of Figure~\ref{fig:en-manual-analysis-cross-ref}}
\label{app:fig:da-manual-analysis-cross-ref}
\end{figure}
\begin{figure}[!h]
\centering
\begin{tikzpicture}
    \node[draw, rounded corners, fill=gray!5, inner sep=10pt, text width=0.95\linewidth, align=left] {
        \textbf{Query (\texttt{serviceloven\_\textsection 82b\_stk2}):}\\
        \emph{Kan en kommunalbestyrelsesbeslutning om at tildele et tidsbegrænset tilskud til individuel hjælp, omsorg, støtte eller optræning -- som kan gives i op til 6 måneder og kun ved vurdering af forbedring eller forebyggelse af forværring -- indbringes for en anden administrativ myndighed?}\\[2ex]
        
        \textbf{Target chunk:}\\
        \emph{Stk.~2. Kommunalbestyrelsens afgørelse efter stk.~1 kan ikke indbringes for anden administrativ myndighed.}\\[2ex]
        
        \textbf{Generated context (Wrong-topic prefix):}\\
        \emph{Dette afsnit omhandler kommunalbestyrelsens afgørelser vedrørende generelle tilbud med aktiverende og forebyggende sigte, herunder fastsættelse af retningslinjer for målgrupper og betaling for tilbuddene.}
    };
\end{tikzpicture}
\caption{Original Danish version of Figure~\ref{fig:en-manual-analysis-wrong-prefix}}
\label{app:fig:da-manual-analysis-wrong-prefix}
\end{figure}
\begin{figure}[!h]
\centering
\begin{tikzpicture}
    \node[draw, rounded corners, fill=gray!5, inner sep=10pt, text width=0.95\linewidth, align=left] {
        \textbf{Query (\texttt{almenboligloven\_\textsection 58b\_stk1}):}\\
        \emph{Kan kommunalbestyrelsen tildele en ledig ældrebolig eller plejehjemsplads, når den har indgået en aftale om, at den almene boligorganisation overtager anvisningen af de almene ældreboliger?}\\[2ex]
        
        \textbf{Target chunk:}\\
        \emph{\textsection~58~b. Ældre og personer med handicap, der har behov for en ældrebolig, en plejehjemsplads eller en beskyttet bolig, optages på en venteliste i bopælskommunen \ldots\ Ledige boliger anvises af kommunalbestyrelsen, \textbf{jf.~dog \textsection~55, stk.~1}, \ldots\ til de personer, som har størst behov for den pågældende bolig og derefter til de personer, som i længst tid har stået på ventelisten.}\\[2ex]
        
        \textbf{Generated context (under-specific prefix):}\\
        \emph{Dette afsnit indgår i kapitel 4, som omhandler udlejning og anvisning af almene boliger, og specifikt i underafsnittet om frit valg af ældreboliger. Afsnittet regulerer optagelse på ventelister for ældre og personer med handicap, der søger ældrebolig, plejehjemsplads eller beskyttet bolig, samt anvisningsrækkefølgen af ledige boliger baseret på behov og ventetid.}
    };
\end{tikzpicture}
\caption{Original Danish version of Figure~\ref{fig:en-manual-analysis-underspecific-prefix}.}
\label{fig:dk-manual-analysis-underspecific-prefix}
\end{figure}
\begin{figure}[!h]
\centering
\begin{tikzpicture}
    \node[draw, rounded corners, fill=gray!5, inner sep=10pt, text width=0.95\linewidth, align=left] {
        \textbf{Query:}\\
        \emph{Betaler staten udgifterne til den gratis rådgivning og vejledning, som ydes til børn og unge med nedsat fysisk eller psykisk funktionsevne samt deres familier?}\\[2ex]
        
        \textbf{Target chunk (Pointer):}\\
        \emph{§ 201. Staten afholder udgifterne til den uvildige konsulentfunktion efter § 163.}\\[2ex]
        
        \textbf{Utilized context chunk (Payload):}\\
        \emph{§ 163. En uvildig konsulentordning yder gratis rådgivning og vejledning i sager om hjælp til børn og unge med nedsat fysisk eller psykisk funktionsniveau og deres familier.}
    };
\end{tikzpicture}
\caption{Original Danish version of Figure~\ref{fig:da-well-formed-query-en}.}
\label{app:fig:da-well-formed-query}
\end{figure}
\FloatBarrier
\section{Impact of the Anthropic-style contextual retrieval on the non-contextual queries}\label{app:anthropic-impact-non-contextual}
\begin{table}[!h]
\centering
\begin{tabular}{@{}lccc@{}}
\toprule
\textbf{Chunk Type} & \textbf{R@1} & \textbf{R@5} & \textbf{R@10} \\ \midrule
Normal Chunks       & 0.575        & 0.861        & 0.921         \\
Augmented Chunks    & 0.611        & 0.896        & 0.945         \\ \bottomrule
\end{tabular}
\caption{Impact of context augmentation on non-contextual queries for the Polish dataset (3,536 queries) using BGE-M3. This table evaluates whether applying LLM-generated contextual augmentation to chunks negatively impacts the retrieval performance of standard, self-contained queries.}
\label{tab:appendix-polish-non-contextual}
\end{table}
\begin{table}[!h]
\centering
\begin{tabular}{@{}lccc@{}}
\toprule
\textbf{Chunk Type} & \textbf{R@1} & \textbf{R@5} & \textbf{R@10} \\ \midrule
Normal Chunks       & 0.557        & 0.885        & 0.944         \\
Augmented Chunks    & 0.603        & 0.909        & 0.961         \\ \bottomrule
\end{tabular}
\caption{Impact of context augmentation on non-contextual queries for the Danish dataset (2,945 queries) using BGE-M3. This table evaluates whether applying LLM-generated contextual augmentation to chunks negatively impacts the retrieval performance of standard, self-contained queries.}
\label{tab:appendix-danish-non-contextual}
\end{table}

\FloatBarrier

\section{Data Sources}\label{app:data-sources}
\begin{figure}[!h]
    \centering
    \includegraphics[width=0.48\textwidth]{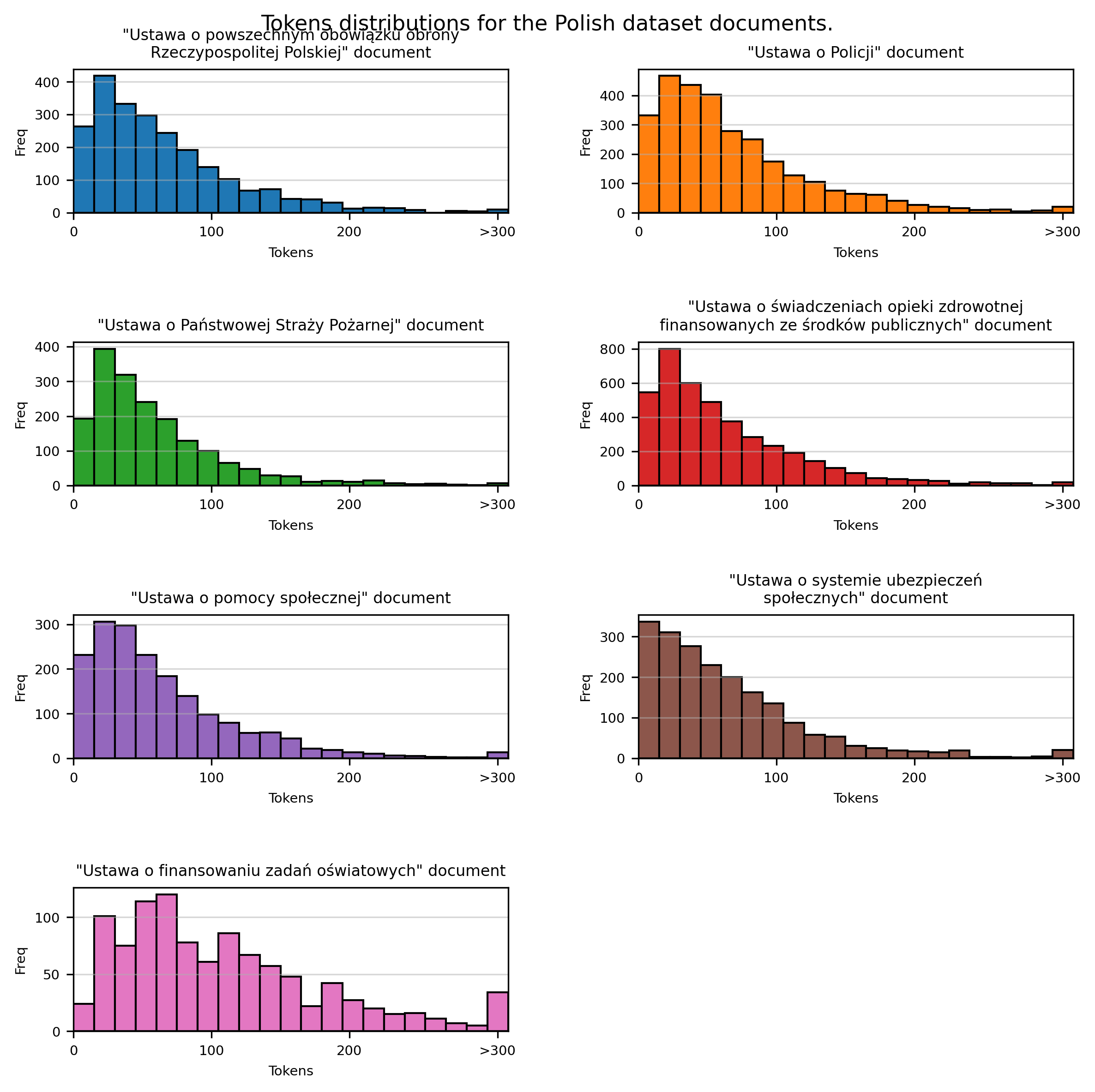}
    \caption{Token distributions of chunks for the Polish dataset, performed using the Byte-Level BPE (BBPE) tokenizer from Qwen.}
    \label{fig:token-distributions-polish}
\end{figure}
\begin{figure}[!h]
    \centering
    \includegraphics[width=0.48\textwidth]{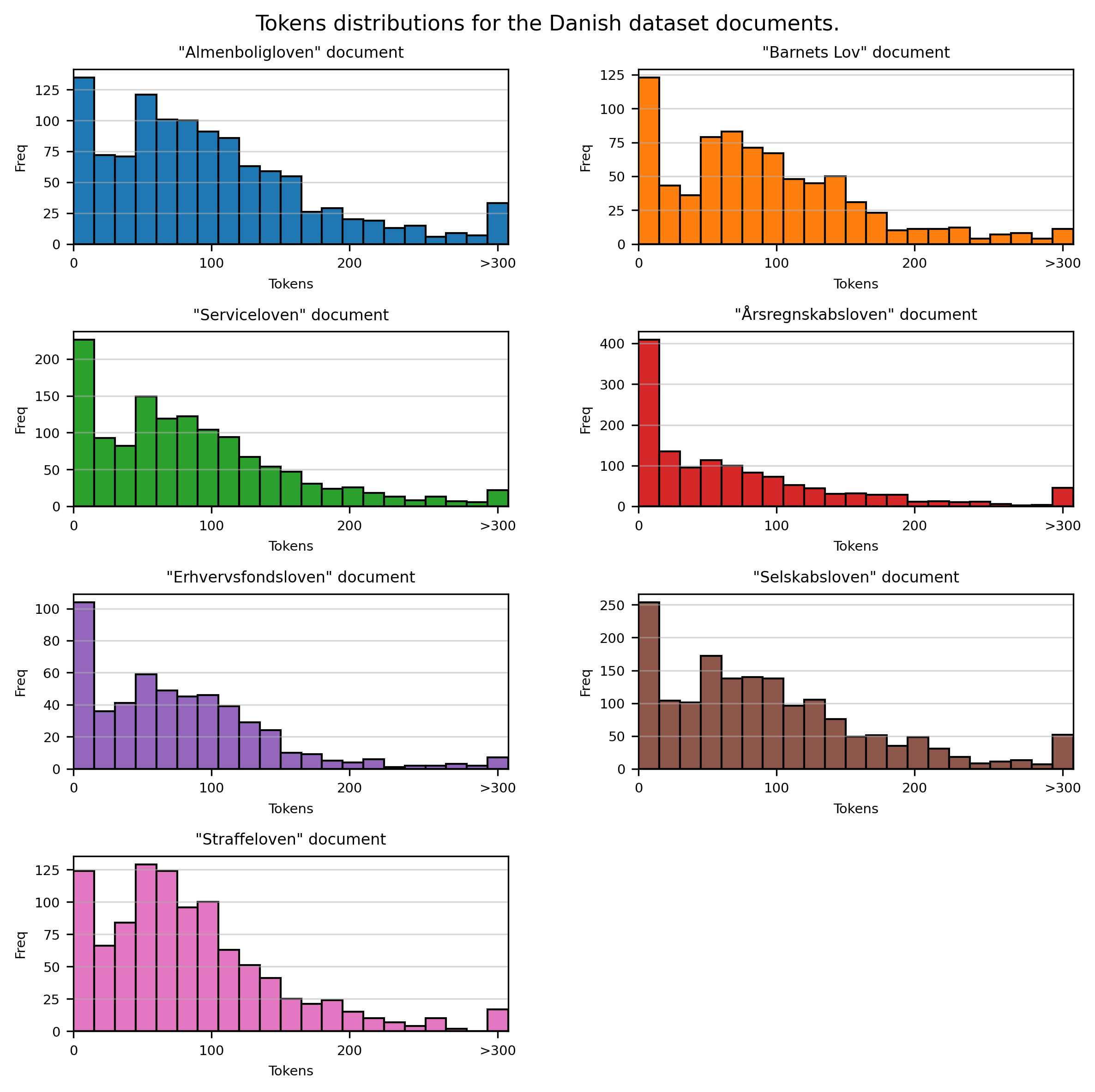}
    \caption{Token distributions of chunks for the Danish dataset, performed using the Byte-Level BPE (BBPE) tokenizer from Qwen.}
    \label{fig:token-distributions-danish}
\end{figure}
\begin{table*}[!h]
    \centering
    \vspace{2mm}
    \begin{tabular}{llcl}
        \hline
        \textbf{Name} & \textbf{\# Chunks} & \textbf{\# Tokens} & \textbf{Description} \\
        \hline
        Almenboligloven & 1131 & 114,716 & Act on Public Housing\\
        Barnets lov & 777 & 71,984 & The Children's Act\\
        Serviceloven & 1325 & 116,108 & The Social Services Act\\
        Årsregnskabsloven & 1327 & 102,365 & The Financial Statements Act\\
        Erhvervsfondsloven & 523 & 40,716 & The Commercial Foundations Act\\
        Selskabsloven & 1647 & 165,983 & The Companies Act\\
        Straffeloven & 1013 & 88,987 & The Danish Penal Code\\
        \hline
    \end{tabular}
    \caption{Overview of chunked Danish documents composing CRAwLeR-DK. These documents were obtained from the Retsinformation website (\url{https://www.retsinformation.dk/}) in April 2026.}
    \label{tab:retsinformation-datasets}
\end{table*}
\begin{table*}[!h]
    \vspace{-20pt}
    \centering
    \vspace{2mm}
    \begin{tabular}{p{6.5cm} c c p{6.5cm}}
        \hline
        \textbf{Name} & \textbf{\# Chunks} & \textbf{\# Tokens} & \textbf{Description} \\
        \hline
        Ustawa o powszechnym obowiązku obrony Rzeczypospolitej Polskiej & 2312 & 153,823 & Act on the Universal Obligation to Defend the Republic of Poland\\
        Ustawa o Policji & 2932 & 204,368 & The Police Act. \\
        Ustawa o Państwowej Straży Pożarnej & 1808 & 107,216 & The State Fire Service Act\\
        Ustawa o świadczeniach opieki zdrowotnej finansowanych ze środków publicznych & 4054 & 259,503 & Act on Publicly Funded Healthcare Services \emph{(two last chapters removed)}\\
        Ustawa o pomocy społecznej & 1820 & 119,436 & The Social Assistance Act\\
        Ustawa o systemie ubezpieczeń społecznych & 2011 & 134,004 & The Social Insurance System Act\\
        Ustawa o finansowaniu zadań oświatowych & 1030 & 114,079 & Act on Financing Educational Tasks\\
        \hline
    \end{tabular}
    \caption{Overview of chunked Polish legal documents composing CRAwLeR-PL. Obtained through ELI API (\url{api.sejm.gov.pl/eli}) in April 2026. }
    \label{tab:polish-legal-datasets}
\end{table*}

\FloatBarrier

\section{Document similarity heatmaps}\label{app:document-similarity-heatmaps}
\begin{figure}[!h]
    \centering
    \includegraphics[width=0.48\textwidth]{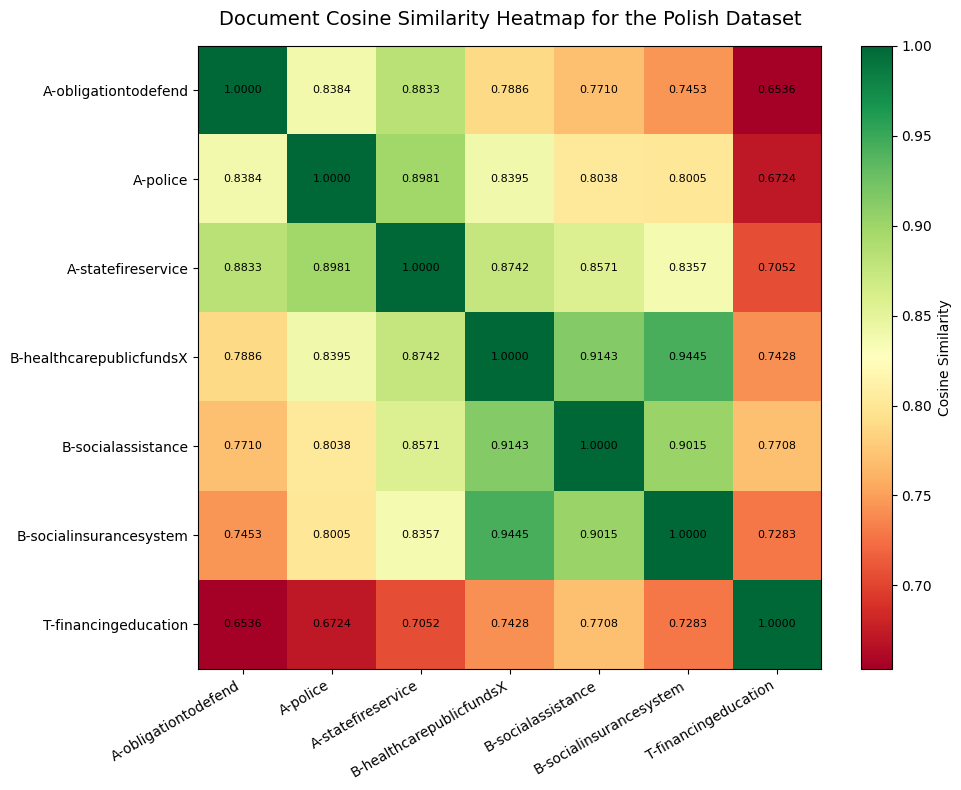}
    \caption{Document Cosine Similarity Heatmap for the CRAwLeR-PL dataset. Computed using \texttt{paraphrase-multilingual-MiniLM-L12-v2} sentence transformer \citep{reimers-2019-sentence-bert}}
    \label{fig:polish-heatmap}
\end{figure}
\begin{figure}[!h]
    \centering
    \includegraphics[width=0.48\textwidth]{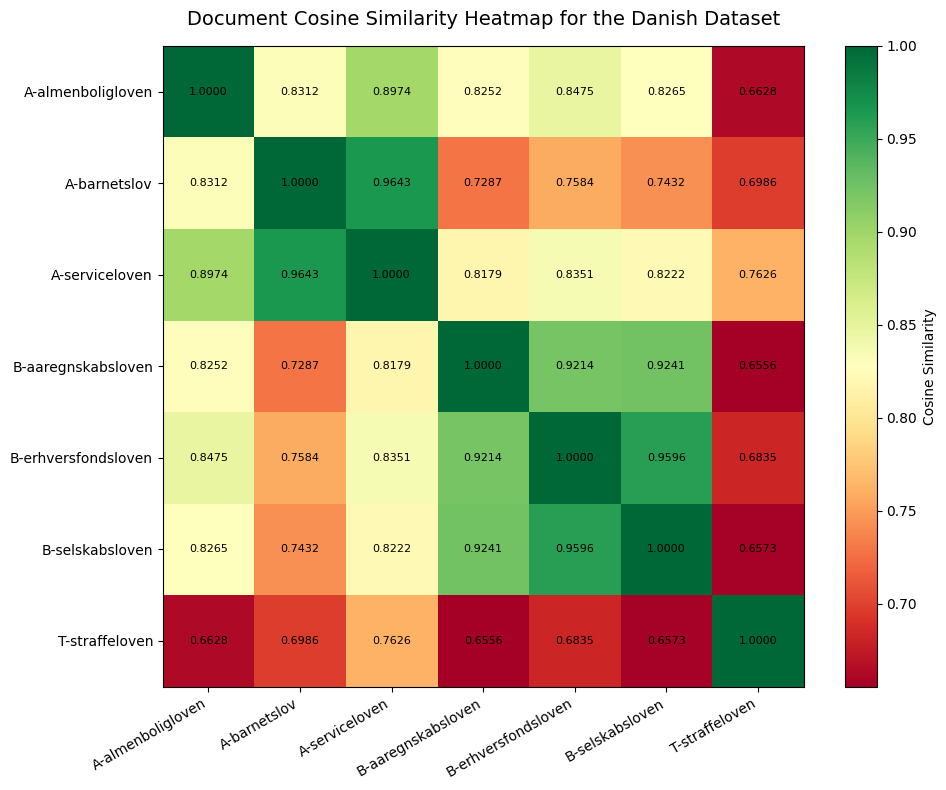}
    \caption{Document Cosine Similarity Heatmap for the CRAwLeR-DK dataset. Computed using \texttt{paraphrase-multilingual-MiniLM-L12-v2} sentence transformer \citep{reimers-2019-sentence-bert}}
    \label{fig:danish-heatmap}
\end{figure}
\FloatBarrier
\section{Code and Artifacts}
The repository is in the abstract's footnote. Inside the repository there is also a link to datasets.

\newpage

\end{document}